\documentclass[twocolumn,aps,prl,nofootinbib]{revtex4-2}
\usepackage{graphicx}
\usepackage{bm}
\usepackage[linktocpage=true,colorlinks=true,urlcolor=blue,linkcolor=red,citecolor = blue]{hyperref}
\usepackage{pgfplots}
\usepackage{amsfonts}
\usepackage{amsmath}
\usepackage{amssymb}
\usepackage{xcolor}
\usepackage{braket}
\usepackage{relsize}
\usetikzlibrary{arrows}
\usepackage{enumerate}
\usepackage{pgfplots}
\usepackage{newfloat}
\usepackage{siunitx}
\usepackage{physics}
\usepackage{mathtools}
\usepackage{comment}
\usepackage{soul}
\usepackage[T1]{fontenc}
\usepackage{times}
\usetikzlibrary {arrows.meta} 
\usepackage{esint}

\newcommand{\printfnsymbol}[1]{%
  \textsuperscript{\@fnsymbol{#1}}%
}
\newcommand{\red}[1]{\textcolor{red}{#1}}

\newcommand{\BS}[1]{\boldsymbol{#1}}
\newcommand{\T}[1]{\text{#1}}

\DeclareFloatingEnvironment[name={Supplementary Figure}]{suppfigure}

\graphicspath{ {Figures/} }

\begin{document}

\title{$g$-factor theory of Si/SiGe quantum dots: spin-valley and giant renormalization effects}
\author{Benjamin D. Woods}
\affiliation{Department of Physics, University of Wisconsin-Madison, Madison, WI 53706, USA}
\author{Merritt P. Losert}
\affiliation{Department of Physics, University of Wisconsin-Madison, Madison, WI 53706, USA}
\author{Robert Joynt}
\affiliation{Department of Physics, University of Wisconsin-Madison, Madison, WI 53706, USA}
\author{Mark Friesen}
\affiliation{Department of Physics, University of Wisconsin-Madison, Madison, WI 53706, USA}

\begin{abstract} 
Understanding the $g$-factor physics of Si/SiGe quantum dots is crucial for realizing high-quality spin qubits.
While previous work has explained some aspects of $g$-factor physics in idealized geometries, the results do not extend to general cases and they miss several important features.  
Here, we construct a theory that gives $g$ in terms of readily computable matrix elements, and can be applied to all Si/SiGe heterostructures of current interest.
As a concrete example, which currently has no $g$-factor understanding, we study the so-called Wiggle Well structure, containing Ge concentration oscillations inside the quantum well.
Here we find a significant renormalization of the $g$-factor compared to conventional Si/SiGe quantum wells.
We also uncover a giant $g$-factor suppression of order $\mathcal{O}(1)$, which arises due to spin-valley coupling, and occurs at locations of low valley splitting.   
Our work therefore opens up new avenues for $g$-factor engineering in Si/SiGe quantum dots.
\end{abstract}

\maketitle

Spin qubits based on Si/SiGe quantum dots are promising candidates for scalable quantum computation \cite{Loss1998,Kloeffel2013,Burkard2023}.
In the simplest scheme, these qubits are defined by the Zeeman splitting $E_z$ of a single-electron spin by a magnetic field $\BS{B}$, which to first approximation is given by $E_z = \mu_B g_0 B$, where $g_0 \approx 2$ is the $g$-factor of a free electron.
However,  spin-orbit coupling (SOC) causes important corrections to this picture~\cite{Winkler2003}.
Although these corrections are small in Si/SiGe, where $|g - g_0| = \mathcal{O}(10^{-3})$ has been observed experimentally~\cite{Kawakami2014,Veldhorst2015,Ferdous2018,Ruskov2018}, it is crucial to understand the origins of the effect, since spin-qubit operations and coherence can depend significantly on local $g$-factor fluctuations~\cite{Veldhorst2014,Jock2018,Liu2021,Wang2024,Losert2024}. 

\begin{figure}[t]
\begin{center}
\includegraphics[width=0.5\textwidth]{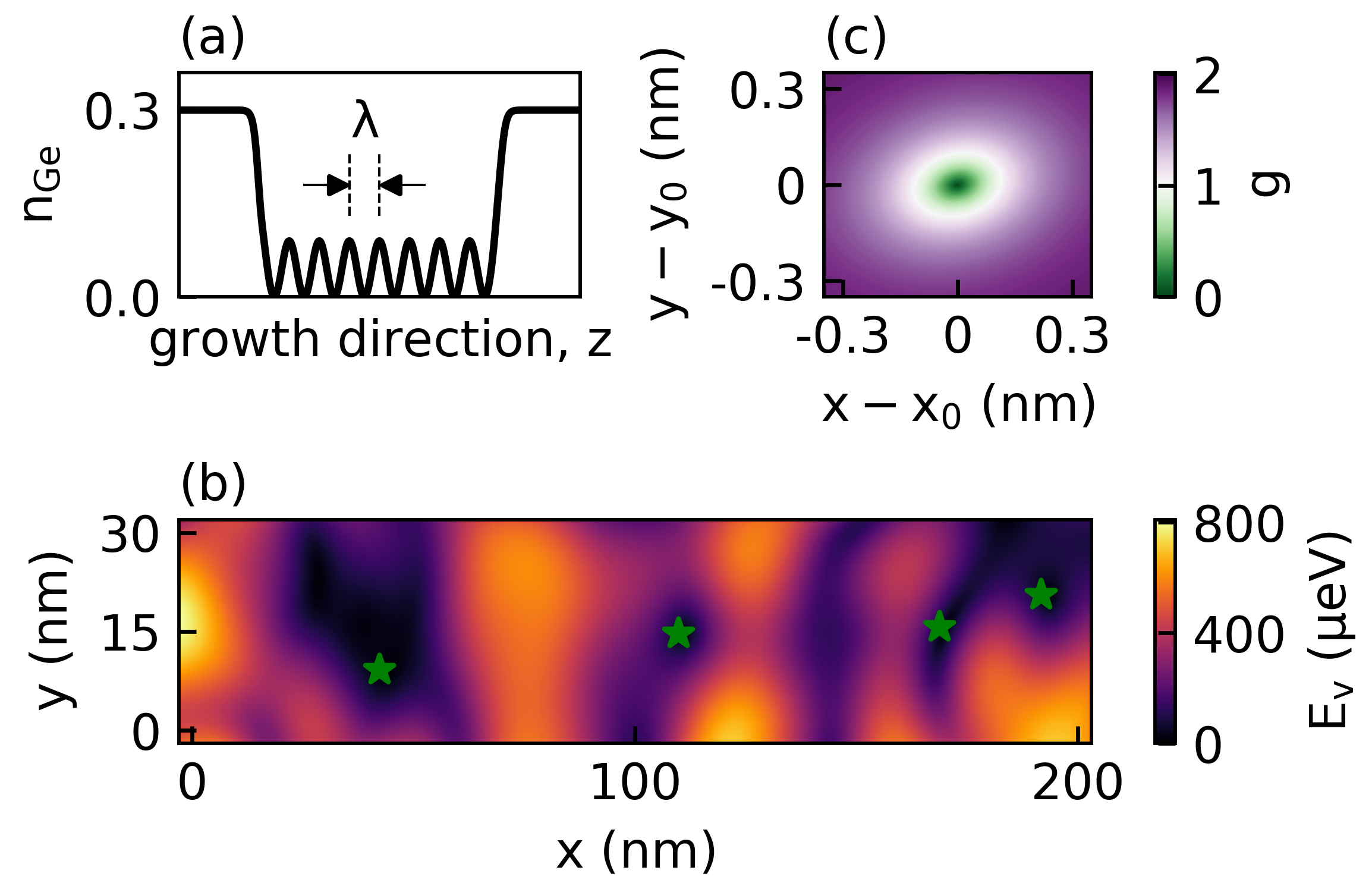}
\end{center}
\vspace{-0.5cm}
\caption{(a) The Ge concentration profile of a Wiggle Well (WW), with Ge concentration oscillations of wavelength $\lambda$, which can significantly renormalize the $g$ factor.
(b) A typical valley-splitting ($E_v$) landscape in a WW, where the green stars indicate the zeros of the valley splitting. 
(c) A spatial map of $g$ near the $E_v = 0$ point at $(x_0, y_0) \!\approx \!(110, 15)~\T{nm}$ in (b).
A giant $g$-factor suppression of order $\mathcal{O}(1)$ occurs in these regions due to the combination of disorder-induced valley-orbit coupling and SOC.
The magnetic field $\BS{B}$ direction has been chosen such that $g = 0$ at the $E_v = 0$ point. 
[See the Supplementary Materials~\cite{SM} for details on generating $E_v$ in (b) and calculating $g$ in (c).]
}
\label{FIG1}
\vspace{-1mm}
\end{figure}

Recent theoretical progress has been made along these lines~\cite{Veldhorst2015,Ruskov2018,Ferdous2018,Ferdous2018b,Nestoklon2008}, for example, by clarifying the dominance of Dresselhaus SOC over Rashba SOC, which explains the observed $B$-field anisotropy of $g$.
However, our understanding is incomplete in other important ways. 
For example, existing results are (i) largely specialized to triangular wells, (ii) rely on a single, vaguely defined Dresselhaus SOC parameter, and (iii) do not fully account for valley physics, as explained below. 
As a result, we show that a giant suppression of the $g$-factor was previously overlooked, and that previous approaches were incapable of describing the $g$-factor physics in the Wiggle Well (WW)~\cite{McJunkin2021}, where the latter refers to Si/SiGe heterostructures with Ge concentration oscillations~\cite{Feng2022,Woods2023a,Woods2024a,Thayil2024}, as illustrated in Fig.~\ref{FIG1}(a).

In this work, we present a general theory for the $g$-factor physics of Si/SiGe quantum dots.
The theory is universal in the sense that $g$ is expressed in terms of readily calculable quantities, such as the SOC coefficients $\beta_{\mu,\nu}$ (defined below), momentum matrix elements $k_z^{\mu,\nu}$, subband energies $\varepsilon_{\nu}$, and valley-orbit matrix elements $\tilde{\Delta}_{\BS{m},\BS{n}}^{\mu,\nu}$, where $\mu$ and $\nu$ are subband indices and $\BS{m}$ and $\BS{n}$ are in-plane orbital indices.
As a concrete example, we apply these results to the WW, where the enhanced SOC from Ge concentration oscillations~\cite{Woods2023a} can cause enhanced $g$-factor fluctuations.
We stress, however, that our results are more broadly applicably to conventional Si/SiGe devices, without Ge concentration oscillations.

Our most interesting finding is a regime in which the $g$-factor variations, $ | g\! - \! g_0|$, are of order $\mathcal{O}(1)$, representing a renormalization that is 3 orders of magnitude larger than previous observations~\cite{Kawakami2014,Veldhorst2015,Ferdous2018,Ruskov2018}.
This dramatic effect is a consequence of spin-valley locking, which results from the combination of enhanced SOC, locally suppressed valley splittings, $E_{v} \lesssim 2~\mu{\T{eV}}$, and valley-orbit coupling, where the latter is caused by random-alloy disorder .
Many such regions of low $E_v$ occur across a typical quantum well \cite{Losert2023}.
In Fig.~\ref{FIG1}(b), we illustrate this for the case of a WW.
Such zeros are readily observed experimentally by means of conveyor-mode shuttling techniques~\cite{Volmer2023} that provide spatial maps of $E_v$. 
If confirmed experimentally, $\mathcal{O}(1)$ suppression of the $g$-factor could be exploited in a variety of settings, such as helping to overcome scaling challenges like frequency crowding~\cite{Heinz2021,Undseth2023,Cifuentes2024} or 
driving singlet-triplet oscillations in double quantum dots~\cite{Jock2018,Liu2021}.
Below, we show that these regions of giant $g$-factor suppression depend sensitively on local random-alloy disorder, but are also robust to random dot displacement caused by charge noise due to their finite spatial extent, as shown in the example $g$-factor map of Fig. \ref{FIG1}(c).
Our work therefore provides interesting new tools for $g$-factor engineering that could be simple to implement in experiments.

This theory also provides insights for $g$-factor physics in the ``normal'' valley splitting regime.
For example, while one could naively expect $g$-factor renormalization and electric dipole spin resonance (EDSR) to depend on the same SOC coefficients, we show here that $g$-factor variations arise from the \textit{inter}subband SOC, while EDSR depends on the \textit{intra}subband SOC~\cite{Rashba2008}. 
This distinction is crucial for understanding the $g$-factor physics of the WW, since the intersubband and intrasubband SOC coefficients depend differently on the Ge oscillation wavelength $\lambda$.
Additionally, we show that the $g$-factor renormalization occurring in a WW in the normal valley splitting regime can be as large as $|g\!-\!g_0| \sim \mathcal{O}(10^{-2})$, which is 1-2 orders of magnitude larger than conventional Si/SiGe quantum wells containing no Ge concentration oscillations~\cite{Kawakami2014,Veldhorst2015,Ferdous2018,Ruskov2018}.

\emph{Model.}---We consider a quantum dot in a Si/SiGe quantum well with an in-plane magnetic field $\BS{B}$ and an arbitrary Ge concentration profile $n_{\T{Ge}}(z)$.
We adopt the gauge $\BS{A} \! = \! A_z(x,y)\hat{\BS{z}}$ for the vector potential, where $A_z(x,y) \! = \! B_x y \! - \! B_y x$. 
The system is modeled by an effective-mass Hamiltonian, where the two valleys 
near the $Z$ point in the Si band structure are treated as a pseudospin-$\frac{1}{2}$ degree of freedom. 
The Hamiltonian can be decomposed into $H = H_v + H_\tau$, where the intravalley term $H_v$ is given by
\begin{equation}
    H_v = \frac{\hbar^2}{2} 
    \left(
    \frac{\hat{\pi}_z^2}
    {m_l} + \frac{\hat{k}_x^2 + \hat{k}_y^2}{m_t}
    \right)
    + V(\BS{r}) 
    + \frac{\mu_B g_0}{2} \BS{B} \cdot \BS{\sigma}, \label{Hv}
\end{equation}
%
$\hat{\pi}_z \! = \! \hat{k}_z + (e/\hbar)A_z(x,y)$,
$\hat{k}_j \! = \! -i\partial_j$, $m_l\! =\! 0.91m_e$ and $m_{t}\! =\!0.19m_e$ \cite{Zwanenburg2013} are the longitudinal and transverse effective masses, respectively, and $\sigma_j$ are Pauli spin matrices, with $j \in \{x,y,z\}$.
The potential is given by 
\begin{equation}
    V(\BS{r}) = 
    V_0 \cos(\frac{2\pi z}{\lambda})
    + V_l(z) 
    + \frac{m_t \omega_t^2}{2} \left(x^2 + y^2\right)
    + V_{\T{dis}}(\BS{r}), \label{POT}
\end{equation}
where the first term, describing a WW, arises from Ge concentration oscillations of wavelength $\lambda$, 
$V_l$ is the quantum well confinement potential arising from the Ge concentration profile $n_{\T{Ge}}(z)$, which also includes the external electric fields but no Ge concentration oscillations or random-alloy fluctuations,
$\hbar \omega_t$ is the orbital splitting of an in-plane harmonic confinement potential representing a quantum dot, 
and $V_{\T{dis}}$ describes the potential fluctuations caused by random-alloy disorder.
The intervalley term $H_{\tau}$ is given by
\begin{equation}
    H_\tau = 
    \left[
    V(\BS{r})
    e^{-i 2k_0 z} + \beta_0 e^{i 2k_1 z}D(\hat{k}_x,\hat{k}_y)
    \right] \tau_- + h.c., \label{Htau}
\end{equation}
where $\tau_j$ are Pauli matrices acting in the $\{+z,-z\}$ valley space, with $j \in \left\{x,y,z\right\}$, $\tau_\pm = (\tau_x \pm i \tau_y)/2$ are valley raising/lowering operators, and ``$h.c.$'' denotes the Hermitian conjugate. 
The first term in Eq.~(\ref{Htau}) describes the normal (i.e. spin-conserving) valley coupling, while the second term is the SOC.
Here, $D(\hat{k}_x,\hat{k}_y)\! =\! \hat{k}_x \sigma_x - \hat{k}_y \sigma_y$ is the Dresselhaus SOC operator whose coupling strength $\beta_0 = 8.2~\T{meV}\cdot \T{nm}$ is determined from a many-band tight-binding theory~\cite{SM}.
The valley and SOC terms both oscillate rapidly, with $2k_0 \approx 0.83 (4\pi/a_{0})$ and $2k_1 = 4\pi/a_{0} - 2k_0$ corresponding to wave vectors connecting valleys in the same or neighboring Brillouin zones, respectively, in the Si band structure. 
Further details regarding the effective-mass model and its connection to a minimal tight-binding model are given in~\cite{SM}.

\emph{Effective Hamiltonian.}---Starting from the effective-mass description of Eqs.~(\ref{Hv}-\ref{Htau}), we now derive a low-energy effective Hamiltonian that captures the universal $g$-factor behavior of Si/SiGe quantum dots.
We begin with two observations.
First, we know from envelope-function theory~\cite{Burt1988} that the eigenstates can be expanded as $\psi(\BS{r}) = \sum_{j} F_j(\BS{r}) U_j(z)$, where $U_j$ are a complete set of periodic functions with periodicity $\lambda$ and $F_j$ are envelope functions with both spin and valley components. 
Provided that the confining potential is smooth on the length scale $\lambda$ and that we choose $U_j$ to be the $\BS{k} = 0$ Bloch functions of the Hamiltonian when only including the periodic component of the potential, we may truncate the series to include only the $j = 0$ term while capturing all effects of the periodic potential to first order in $V_0$ \cite{Woods2024a}.
Note that $U_0(z) \!= \!1 \!+\! (V_0/E_\lambda) \cos(G_\lambda z)$, with $G_\lambda = 2\pi/\lambda$ and $E_\lambda = \hbar^2 G_{\lambda}^2/(2 m_l)$ being the wavevector and energy scale of the Ge concentration oscillations, respectively \cite{Woods2024a}.
Second, the Zeeman coupling, the vector-potential coupling, the intervalley coupling, and the alloy-disorder terms in Eqs.~(\ref{Hv}-\ref{Htau}) are all small compared to the kinetic energy and confinement terms.
Therefore, the low-energy eigenstates in the absence of these perturbations serve as a good basis set for perturbation theory.
These basis functions are given by $\ket{\nu,n_x,n_y,\tau,\sigma}$, where
$\nu$ and $\BS{n}\!=\! (n_x,n_y)$ are spatial orbital indices specified below, and $\tau \in \{+z,-z\}$ and $ \sigma \in \{\uparrow,\downarrow\}$ denote the valley and spin, respectively. 
Here, $\bra{\BS{r}}\ket{\nu,n_x,n_y} = \varphi_{\nu}(z) \chi_{n_x,n_y}(x,y)$,
where $\chi_{n_x,n_y}$ is an in-plane harmonic-oscillator orbital with energy $\hbar\omega_t(n_x\! +\! n_y\!+\!1)$ coming from the harmonic confinement $(m_t \omega_t^2/2)(x^2\!+\!y^2)$, and $\varphi_{\nu}(z) = \varphi_{\nu}^{\T{env}}(z)U_0(z)$ is a subband wavefunction with $\varphi_{\nu}^{\T{env}}$ being the subband wavefunction \textit{envelope} satisfying
\begin{equation}
    \left[\frac{\hbar^2 \hat{k}_z^2}{2m_{t}} + V_l(z)\right]\varphi_{\nu}^{\T{env}}(z) = \varepsilon_\nu \varphi_{\nu}^{\T{env}}(z), \label{SubbandEq}
\end{equation}
where $\varepsilon_{\nu}$ is the subband energy.

An effective Hamiltonian for the lowest-energy orbital $(\nu,n_x,n_y) = (0,0,0)$ may now be derived 
using a Schrieffer-Wolff transformation \cite{Winkler2003,Luttinger1955}.
Including all terms up to second-order yields the effective Hamiltonian~\cite{SM}
\begin{equation}
\begin{split}
    H_{\T{eff}} =& 
    ~\Delta \tau_- + \Delta^* \tau_+ 
    + \frac{\mu_B g_0}{2} \left(\BS{B} + \BS{B}_\text{sv} \tau_z\right) \cdot \BS{\sigma}
    \\
    &+ \frac{\mu_B}{2}\left(g_{\tau} \tau_- + g_{\tau}^* \tau_+\right)\left(B_y \sigma_x + B_x \sigma_y\right), \label{Heff}
\end{split}
\end{equation}
where $\Delta\! \approx \! \mel{0,0,0}{V(\BS{r})e^{-i2k_0z}}{0,0,0}$ is the ``normal''
intervalley coupling, which is related to the (nominal\footnote{There is a correction to the valley splitting from $\BS{B}_\text{sv}$. However, $E_{v} \approx 2|\Delta|$ unless $|\Delta|$ is very small, as discussed below.}) valley splitting by $E_v = 2|\Delta|$.
The second line in Eq.~(\ref{Heff}) contains an intervalley Zeeman coupling term, where $g_{\tau}$ is given by
\begin{equation}
    g_{\tau} =  
    \frac{2e \hbar }{\mu_B m_l}
    \sum_{\nu >0}\frac{(ik_z^{0,\nu}) \beta_{\nu,0}}{\varepsilon_{\nu} - \varepsilon_{0} + \hbar \omega_t}  , \label{gtau0}
\end{equation}
and $\beta_{\mu,\nu} \! = \! \beta_0\mel{\mu}{e^{i2k_1z}}{\nu}$ and $k_z^{\mu,\nu} = \mel{\mu}{\hat{k}_z}{\nu}$ are SOC and momentum matrix elements, respectively.
The intervalley Zeeman term arises from second-order virtual processes involving spin-orbit coupling and coupling to the magnetic vector potential $\BS{A}$, as illustrated in Fig. \ref{FIGS2} of~\cite{SM}.
We note that intra-subband scattering processes ($\nu = 0$) do not contribute to Eq.~(\ref{gtau0}), because the momentum expectation value of any subband wavefunction must vanish: $k_z^{\nu,\nu} = 0$.

The enhanced SOC of the WW gives an important new term in Eq.~(\ref{Heff}), proportional to $\tau_z \BS{\sigma}$.
We refer to the coupling constant $\BS{B}_\text{sv}$ as a spin-valley ``field'' because of the way it couples to the spin.
Importantly, it gives a nonvanishing spin splitting in both the $\pm z$-valleys (with opposite quantization axes due to the $\tau_z$ factor) even in the limit $B = 0$.
Here, the $x$ component of $\BS{B}_\text{sv}$ is given by
\begin{equation}
    B_\text{sv}^{x} = \frac{-4}{\mu_B g_0} \sqrt{\frac{ m_t \omega_t}{2\hbar}} \sum_{\nu \geq 0} 
   \frac{\Im[
   \beta_{0,\nu}^* \tilde{\Delta}_{10,00}^{\nu,0}
   ]}
   {
   {\varepsilon}_{\nu} - \varepsilon_{0} + \hbar \omega_t
   },\label{Bsvx}
\end{equation}
while $B_\text{sv}^{y}$ is obtained by substituting $\tilde{\Delta}_{10,00}^{\nu,0} \rightarrow - \tilde{\Delta}_{01,00}^{\nu,0}$.
This spin-valley field arises from second-order processes involving SOC (the $\beta_{0,\nu}^*$ factors) and the valley-orbit coupling, defined as $\tilde{\Delta}_{\BS{m},\BS{n}}^{\mu,\nu} \! =\! \mel{\mu,m_x,m_y}{V(\BS{r})e^{-i2k_0z}}{\nu,n_x,n_y}$, where we note that $\Delta\!\approx\! \tilde{\Delta}_{\BS{0},\BS{0}}^{0,0}$ (see \cite{SM}).
In contrast to $g_{\tau}$ in Eq.~(\ref{gtau0}), the sum in Eq.~(\ref{Bsvx}) includes the $\nu = 0$ subband. 
Indeed, since the subband energy splitting is much larger than the in-plane orbital splitting ($\varepsilon_{\nu \neq 0} \! - \! \varepsilon_{0} \gg \hbar \omega_t$), the sum is dominated by the intrasubband ($\nu = 0$) term. 
We note that the valley-orbit coupling $\tilde{\Delta}^{\nu,0}_{10,00}$ in Eq.~(\ref{Bsvx}) arises from alloy disorder, and is therefore larger, on average, when the wave function overlaps with regions of higher Ge concentration \cite{Losert2023}.
The complex phase of $\tilde{\Delta}_{10,00}^{\nu,0}$ is randomized by alloy disorder, leading to random orientations of $\BS{B}_\text{sv}$ in the $x$-$y$ plane.
We also note here that the spin-valley term preserves time-reversal symmetry, ensuring Kramer's degeneracy when $\BS{B} = 0$.

We now wish to compute the $g$ factor, defined as $g \!= \! (E_1 \! - \! E_0)/(\mu_B B)$ when $B \! \rightarrow \! 0$, for both the ground and excited valley states.
Here $E_0$ and $E_1$ are the energies of the spin states in each of the valley subspaces.
(The $g$ factors can differ in these subspaces.)
Diagonalizing Eq.~(\ref{Heff}) in the limit $B\rightarrow 0$, we find
\begin{equation}
    \begin{split}
    g_\pm =  
    \Big(
    &\left[ 
    g_0 \cos\theta_B^\prime 
    \mp |g_\tau| \zeta 
    \cos\phi_v^\prime \sin(\theta_B + \theta_\text{sv})
    \right]^2
    \\
    +&\left[
    g_0 \zeta \sin \theta_B^\prime
    \mp |g_\tau| \cos \phi_v^\prime \cos(\theta_B + \theta_\text{sv})
    \right]^2 
    \Big)^{1/2},
    \end{split}
    \label{geff1}
\end{equation}
where the $g_+$ and $g_-$ refer to the ground and excited valley $g$-factors, respectively, $\zeta = |\Delta|/\sqrt{|\Delta|^2 + \varepsilon_\text{sv}^2/4}$, $\varepsilon_\text{sv} = \mu_B g_0 B_\text{sv}$ is the spin-valley energy, $\theta_B$ and $\theta_\text{sv}$ are the angular orientations of $\BS{B}$ and $\BS{B}_\text{sv}$ with respect to the $[100]$ crystallographic axis, and we define the relative angles $\theta_B^\prime = \theta_B - \theta_\text{sv}$ and $\phi_v^\prime = \phi_v - \phi_g$, where $\phi_v$ and $\phi_g$ are the complex phases of $\Delta$ and $g_{\tau}$.
Note that we have ignored a term in Eq.~(\ref{geff1}) proportional to $|g_\tau|^2$ since $|g_{\tau}| \ll g_0$.
Importantly, Eqs.~(\ref{gtau0})-(\ref{geff1}) describe results for $g$ that are not limited to flat quantum wells with constant electric fields, in contrast to  Refs.~\cite{Veldhorst2015,Ruskov2018,Ferdous2018,Ferdous2018b}.
This is crucial when considering a range of important geometries, including the WW, and quantum wells containing disorder or interfaces with nonzero width.

As a specific example, we now consider the interesting case of a long-period WW, with $\lambda \approx \pi/k_1 \approx 1.57~\text{nm}$, which is known to enhance SOC~\cite{Woods2023a}.
It is well understood that super-sharp quantum-well interfaces capable of deterministically enhancing the valley splitting are very difficult to grow in the laboratory~\cite{Losert2023}.
In order to focus on the WW physics here, we remove sharp interfaces from the picture by adopting a soft quantum-well confinement potential, given by $V_l(z) = m_l \omega_z^2 z^2/2$, where $\hbar \omega_z$ is a harmonic subband spacing.
We emphasize however that the current approach can also be applied to conventional heterostructures.
Below, we analyze Eq.~(\ref{geff1}) in the normal and low-valley-splitting regimes, corresponding to $\zeta \approx 1$ and $\zeta \ll 1$, respectively.

\begin{figure}[t]
\begin{center}
\includegraphics[width=0.48\textwidth]{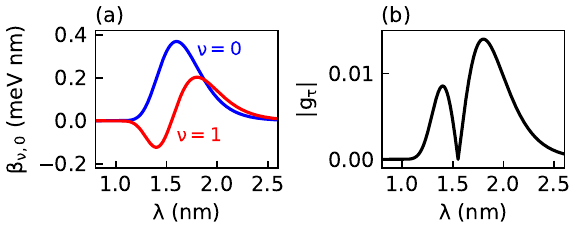}
\end{center}
\vspace{-0.5cm}
\caption{(a) The SOC coefficients $\beta_{\nu,0}$ of a WW, plotted as a function of the Ge oscillation wavelength $\lambda$.
The quantum well confinement energy is $\hbar \omega_z \! = \! 20~\T{meV}$ and the WW has an average Ge concentration of $\bar{n}_{\T{Ge}} = 0.05$, corresponding to an oscillating potential amplitude of $V_0 = 30~\T{meV}$. 
The intrasubband ($\beta_{0,0}$) and intersubband ($\beta_{1,0}$) SOC coefficients are strongly enhanced near the long-period WW wavelength $\lambda = \pi/k_1 = 1.57~\T{nm}$, although $\beta_{1,0}$ has a node at this value.
(b) The intervalley Zeeman renormalization parameter $|g_{\tau}|$ for the same quantum well as (a), with quantum dot confinement energy $\hbar \omega_t = 2~\T{meV}$.
Here, $|g_{\tau}|$ exhibits a similar dependence on $\lambda$ as $|\beta_{1,0}|$.
}
\label{FIG4}
\vspace{-1mm}
\end{figure}

\emph{Normal-valley-splitting regime.}---The most common situation for a Si/SiGe quantum well is for the valley splitting to be much larger than the spin-valley energy, $|\Delta| \gg \varepsilon_\text{sv}$.
We call this the ``normal'' valley splitting regime, for which we can simply ignore the spin-valley field by setting $\zeta = 1$.
Equation~(\ref{geff1}) then reduces to
\begin{equation}
    g_\pm \approx g_{0} 
    \mp |g_{\tau}| \cos\left(\phi_{v} - \phi_g\right) \sin(2 \theta_B ). \label{geff} 
\end{equation}
Here, $g_\pm$ is anisotropic with respect to the magnetic field orientation, contained in the $\sin(2\theta_B)$ factor. 
This anisotropy agrees with previous theoretical work~\cite{Veldhorst2015,Ruskov2018,Ferdous2018,Ferdous2018b,Cifuentes2024} and arises from the breaking of axial rotational symmetry by the Dresselhaus SOC.
The overall magnitude of the $g$-factor renormalization is set by $|g_\tau|$.
To compute $g_\tau$ from Eq.~(\ref{gtau0}), we first need to calculate the longitudinal momentum $k_z^{0,\nu}$ and SOC $\beta_{\nu,0}$ matrix elements. 
For harmonic-oscillator wave functions, we obtain the simple result that $k_{z}^{\mu,\nu} = i\sqrt{\hbar/(2m_l \omega_z)}(\sqrt{\mu}\delta_{\mu,\nu + 1} - \sqrt{\nu}\delta_{\mu,\nu - 1})$, showing that only the $\nu = 1$ subband contributes to the $g$-factor renormalization. 
The SOC coefficients are given by
\begin{align}
    \beta_{0,0} =& ~
    \beta_0
    e^{-(2 k_1 \ell_z)^2} + 
    \frac{\beta_0 V_0}{E_{\lambda}}
    e^{-\kappa^2 \ell_z^2}
    ,  \label{beta00}\\
    \beta_{1,0} =&~ 
    \beta_0 \ell_z 
    2 k_1  e^{-(2 k_1 \ell_z)^2} 
    + \frac{\beta_0 V_0}{E_{\lambda}}
    \kappa \ell_z
    e^{-\kappa^2 \ell_z^2}
    , \label{beta10}
\end{align} 
where $\ell_z \! = \! \sqrt{\hbar/(2m_l \omega_z)}$ and $\kappa \! = \! 2k_1 \! - \! G_\lambda$.
We note that the first terms in each of these equations correspond to conventional results, with no WWs ($V_0=0$), which are strongly suppressed in the harmonic well, due to the large argument inside the exponential functions.
On the other hand, when wiggles are present ($V_0>0$) and the condition $\lambda = \pi/k_1 \approx 1.57~\text{nm}$ is fulfilled, such that $\kappa\approx 0$, the $\beta_{\nu,0}$ terms are no longer exponentially suppressed, as shown in Fig.~\ref{FIG4}(a).
Here, we note that $\beta_{0,0}$ and $\beta_{1,0}$ exhibit different behaviors, with the former being maximized at $\lambda = \pi/k_1$, while the latter exhibits a node at this value.
(See~\cite{SM} for explanatory details.)

In Fig.~\ref{FIG4}(b) we show results for the renormalization factor $|g_\tau|$, rather than $g$ itself, because the latter is modified by the randomized valley phase $\phi_v$ in Eq. (\ref{geff}). 
Far away from the WW condition ($\lambda = 1.57~\T{nm}$), $|g_\tau|$ is exponentially suppressed because the 
harmonic confinement lacks a sharp interface. 
Near $\lambda = 1.57~\T{nm}$, however, $|g_\tau|$ is on order of $10^{-2}$, representing an order of magnitude enhancement compared to conventional Si/SiGe quantum wells~\cite{Kawakami2014,Veldhorst2015,Ferdous2018,Ruskov2018}.
Notably, $|g_{\tau}|$ displays  the same dependence on $\lambda$ as $|\beta_{1,0}|$, with a node at $\lambda = 1.57~\T{nm}$. 
This working point could be useful when performing EDSR-based quantum gates, because the intrasubband SOC is maximized, while $g_\tau$ vanishes, leaving a $g$-factor that is highly stable against charge noise.

\emph{Giant suppression of $g$ (low-valley-splitting regime).}---The presence of Ge in the quantum well increases the valley splitting on average, but also enhances valley splitting variability. 
This occasionally results in locations with small $|\Delta|$~\cite{Losert2023}, as shown in Fig. \ref{FIG1}(b).
At such locations, we may have $|\Delta| \ll \varepsilon_\text{sv}$, such that $\zeta \rightarrow 0$, due to enhanced spin-valley coupling in the WW.
Equation~(\ref{geff1}) then reduces to
\begin{equation}
    g \approx g_0 |\cos \theta_B^\prime| . \label{geff2}
\end{equation}
(Here, we have ignored corrections proportional to $g_{\tau}$, since they are relatively very weak.
Note also that $g=g_+=g_-$ in this case.)
The resulting suppression of $g$ is of order $g_0$, which is so large that it has no precedent in recent Si/SiGe experiments~\cite{Kawakami2014,Veldhorst2015,Ferdous2018,Ruskov2018}. 
In fact, for the special field orientation for which $\BS{B}$ and $\BS{B}_\text{sv}$ are orthogonal ($\theta'_B=\pi/2$), $g$ can even vanish.

We can understand the dramatic suppression of $g$ as follows. 
When the spin-valley coupling dominates in Eq.~(\ref{Heff}), $\BS{B}_\text{sv}$ becomes the quantization axis.
To better understand the effect of this, we rotate the spin so that $\BS{B}_\text{sv}$ points along the $z$ axis.
Again ignoring terms proportional to $g_\tau$, we have 
\begin{equation}
    \begin{split}
    H_{\T{eff}}^{\prime} = &~ 
    \frac{\varepsilon_\text{sv}}{2}  \tau_z \sigma_z 
    + |\Delta| \tau_x     + \frac{E_z}{2} \left(
    \cos\theta_{B}^\prime \sigma_z + \sin \theta_{B}^\prime \sigma_x
    \right),
    \end{split} \label{HeffPrime}
\end{equation}
where $E_z = \mu_B g_0 B$ is the unrenormalized Zeeman energy.
Note that we have rotated the valley and spin about the $\tau_z$ and $\sigma_z$ axes, respectively, to remove terms proportional $\tau_y$ and $\sigma_y$.
In the limit $|\Delta| \ll \varepsilon_\text{sv}$, and further assuming small magnetic fields for which $E_z \ll \varepsilon_\text{sv}$, the first term in Eq.~(\ref{HeffPrime}) dominates.
The resulting eigenstates are then well-defined spin-valley states, and the spins and valleys are said to be locked, with opposite spin orientations for the two $\pm z$-valley states.
This is similar to spin-valley locking in certain 2D materials with strong SOC, such as 2D transition metal dichalcogenides~\cite{Tao2019}.
Now, 
applying first-order perturbation theory by projecting onto the two lowest-energy states,
we see that only the magnetic field component parallel to $\BS{B}_\text{sv}$ contributes to the spin splitting.
Thus, spin-valley locking can be understood as the source of the giant $g$-factor renormalization in Eq.~(\ref{geff2}), because it decouples the quantization axis from the magnetic field axis.

\begin{figure}[t]
\begin{center}
\includegraphics[width=0.5\textwidth]{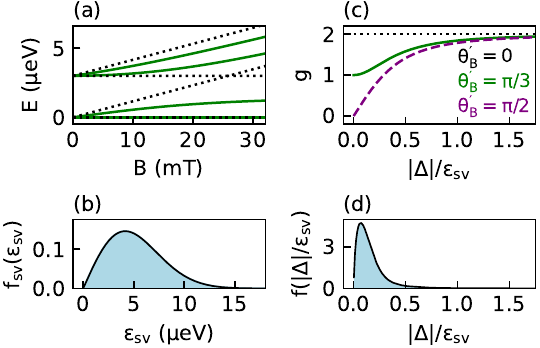}
\end{center}
\vspace{-0.5cm}
\caption{
(a) The energy spectrum of the lowest spin-valley states obtained from Eq.~(\ref{HeffPrime}), as a function of $B$, for the model parameters $\varepsilon_\text{sv} = 2.5~\mu\T{eV}$, $\Delta = 0$, and either $\theta_B^\prime = 0$ (dotted black line) or $\theta_B^\prime = \pi/3$ (solid green line). 
(b) A typical $\varepsilon_\text{sv}$ distribution for a long-period WW. 
Typical $\varepsilon_\text{sv}$ values are in the range of a few $\mu\T{eV}$, implying that a giant $g$-factor suppression can occur for fields of up to tens of mT.
(See the main text for device parameters.)
(c) $g$ as a function of $|\Delta|/\varepsilon_\text{sv}$, for the indicated $\theta_B^\prime$ values. 
A giant $g$-factor suppression is observed in the range $|\Delta|/\varepsilon_\text{sv} =E_v/2\varepsilon_\text{sv} \lesssim 1$.
(d) Typical distribution of $|\Delta|/\varepsilon_\text{sv}$ in the presence of charge noise (see the main text for details), for the same WW as (b).
Here, the dot is tuned to a point where $\Delta=0$ in the absence of charge noise. 
The distribution is concentrated in the range $|\Delta|/\varepsilon_\text{sv} \ll 1$, indicating that charge noise does not wash out the $g$-factor suppression.
}
\label{FIG5}
\vspace{-1mm}
\end{figure}

For higher magnetic fields, the first-order perturbation theory breaks down, and we must include the $(E_z/2) \sin \theta^\prime_B\sigma_x$ term in the analysis. Diagonalizing Eq.~(\ref{HeffPrime}) then yields a non-linear magnetic field dependence, as illustrated in Fig.~\ref{FIG5}(a), where we have chosen $\varepsilon_\text{sv} \! = \! 2.5~\mu\T{eV}$ (or $B_\text{sv} \approx 20~\T{mT}$), $\Delta=0$, and $\theta_B^\prime\! = \! \pi/3$, yielding $g\! = \! g_0/2$ as $B \rightarrow 0$.
In this case, the spectrum splits into well-defined lower and upper valley-state manifolds, which are separated by the spin-valley energy $\varepsilon_\text{sv}$ at low fields, rather than the conventional valley splitting.
When $B \gtrsim B_\text{sv} $, we observe significant deviations from linear spin splitting, indicating that the $g$ factor is no longer well defined.

The magnetic field $B$ at which the system deviates from linear splitting is clearly controlled by the the strength of the spin-valley field $B_\T{sv}$ (or equivalently $\varepsilon_\T{sv}$).
Therefore, it is important to understand the distribution of $\varepsilon_\T{sv}$, which shows spatial fluctuations due to random-alloy disorder. 
In~ \cite{SM}, we show that $\varepsilon_\text{sv}$ follows a Rayleigh distribution, similar to $E_v$.
Importantly, the scale parameter of the Rayleigh distribution is $\sigma_\T{sv}\! \propto\! |\beta_{0,0}|\sqrt{\bar{n}_{\T{Ge}}}$. 
Therefore, the WW, with its enhanced SOC and finite Ge density in the quantum well region, can be expected to typically yield large $\varepsilon_\T{sv}$.
Figure~\ref{FIG5}(b) shows a distribution of $\varepsilon_\text{sv}$ for a WW with Ge oscillations of wavelength $\lambda = 1.57~\T{nm}$ and an average Ge concentration of $\bar{n}_{\T{Ge}} = 0.05$, corresponding to a potential oscillation amplitude of $V_0 = 30~\T{meV}$.
This yields typical $\varepsilon_\text{sv}$ values of several $\mu\T{eV}$ (corresponding to $B_\text{sv}>20$~mT), indicating that the linear-regime of the giant $g$-factor suppression should be readily observed in magnetic fields of up to tens of mT.

In addition to low magnetic fields, giant $g$-factor renormalization also requires low $|\Delta|$ values. 
Indeed, we find the exact $g$-factor of the effective Hamiltonian in Eq. (\ref{Heff}) to be 
\begin{equation}
    g \approx \frac{\sqrt{4|\Delta|^2 + \varepsilon_\T{sv}^2 \cos^2 \theta_B^\prime}}{\tilde{E}_v} g_0, \label{gFactor}
\end{equation}
where $\tilde{E}_v = \sqrt{4 |\Delta|^2 + \varepsilon_\T{sv}^2}$ is the generalized valley splitting in the presence of the spin-valley field.
We plot $g$ in Fig. \ref{FIG5}(c) as a function of $|\Delta|/\varepsilon_\T{sv}$ for three values of $\theta_{B}^\prime$ (the relative angle between $\BS{B}$ and $\BS{B}_\text{sv}$).
Here, $g$ increases monotonically with increasing $|\Delta|/\varepsilon_{sv}$, where $g \rightarrow g_0$ is recovered in the ``normal'' regime limit of $|\Delta|/\varepsilon_{sv} \rightarrow \infty$ independent of $\theta_{B}^\prime$.
We see that the giant $g$-factor suppression is limited to the range $|\Delta|/\varepsilon_\T{sv}=E_v/2\varepsilon_\T{sv} \lesssim 1$. 
Interestingly, we find that the $g$-factor has a first-order sweet spot at $|\Delta|/\varepsilon_\T{sv} = 0$ with respect to fluctuations in $\Delta$, except in the extreme case of $\theta_{B}^\prime = 0$, where $g$ vanishes.

Given that $E_v$ naturally varies in space, as shown in Fig.~\ref{FIG1}(b), 
the analysis of the previous paragraph suggests an important practical question:
if a dot is tuned to the special working point where $\Delta$ and $g$ are minimized, do electric-field (``charge'') fluctuations shift the dot position enough to destroy the $g$-factor suppression coming from the spin-valley field?
Fortunately, the answer is no in the case of a WW, due to its enhanced spin-orbit coupling.
To support this claim, we assume the dot is initially tuned to a point where $\Delta=0$, in the absence of charge noise.
We then introduce charge noise as a random in-plane electric field $\BS{E}$.
As we show in~\cite{SM}, this displacement field causes 
a change in the ratio between the valley coupling and spin-valley energy, given by
\begin{equation}
    \frac{|\Delta|}{\varepsilon_\text{sv}} = 
    \frac{\hbar^2|\BS{E}|}{2 m_t |\beta_{0,0}| \hbar \omega_t}  
    \Xi(\tilde{\Delta}) ,
    \label{Ratio}
\end{equation}
where $\Xi$ is a function of the various valley-orbit matrix elements $\tilde{\Delta}_{\BS{m},\BS{n}}^{\mu,\nu}$.
Remarkably, we show in~\cite{SM} that $\Xi$ follows a universal distribution that is independent of the details of the quantum well.
Therefore, the distribution of $|\Delta|/\varepsilon_\T{sv}$ only depends on the strength of the charge noise, the SOC coefficient $\beta_{0,0}$, and the orbital splitting $\hbar \omega_t$.
To determine typical values of $|\Delta|/\varepsilon_\text{sv}$, we assume that each component of $\BS{E}$ is normally distributed, with a mean of zero and a standard deviation of $\sigma_E = 20~\mu\T{V/nm}$.
For a double dot geometry with dots separated by $d = 100~\T{nm}$, this corresponds to detuning fluctuations with a standard deviation of $\sigma_{\varepsilon} = 20~\mu\T{eV}$. 
(This is a relatively large charge noise~\cite{Kranz2020}, so we are in no danger of making a conservative estimate.)
In Fig.~\ref{FIG5}(d), we show the resulting distribution of $|\Delta|/\varepsilon_\text{sv}$ values, for a WW with the same parameters as Figs.~\ref{FIG5}(b) and \ref{FIG5}(c), and $\hbar \omega_t = 1~\T{meV}$.
Importantly, the distribution is strongly concentrated in the range of $|\Delta|/\varepsilon_\text{sv} \ll 1$.
This confirms the claim that charge noise does not wash out the desired $g$-factor suppression.

\emph{Conclusion.}---We have developed a theory of $g$-factor physics in Si/SiGe quantum dots that expresses $g$ in terms of readily computable matrix elements and applies to all heterostructures of current interest. 
We use this understanding to reproduce existing results~\cite{Veldhorst2015,Ruskov2018,Ferdous2018,Ferdous2018b,Nestoklon2008}, while also obtaining solutions that were previously unknown. 
In particular, we focus on the case of a long-period WW~\cite{McJunkin2021}, where we observe a $g$-factor renormalization approximately ten times larger than conventional, pure-Si quantum wells.
We also uncover a giant suppression of the $g$ factor at spatially varying locations where $E_v, E_z \lesssim \varepsilon_\text{sv}$, which is of great interest for qubit gate operations, because it allows magnetic field couplings to be suppressed or extinguished.
Importantly, we have shown that charge-noise displacements of a dot away from such interesting working points do not significantly affect their operation.

Our results are also of fundamental interest for all types of Si/SiGe quantum wells, even those without Ge concentration oscillations.
For example, the underlying connection between the $g$-factor physics and valleys is given in Eq.~(\ref{geff}), which clarifies the relationship between $g$, the valley phase $\phi_v$, and the phase $\phi_g$ of the valley-dependent  $g$-factor renormalization $g_{\tau}$.
Although Refs.~\cite{Veldhorst2015,Ruskov2018} include a phase parameter similar to $\phi_g$, it was introduced phenomenologically.
In contrast, $\phi_g$ is derived rigorously in Eq.~(\ref{gtau0}), where it arises from the intersubband SOC.
This has important implications for the valley-dependent $g$-factor renormalization $g_{\tau}$, which is found to be robust against alloy-disorder fluctuations, due to the large energy spacing between subbands as compared to the energy scale of the alloy disorder. 
In contrast, the valley phase $\phi_v$ is mainly determined by the locally varying alloy disorder~\cite{Losert2023}, as reflected in local fluctuations of the giant $g$-factor in WWs.
We will make use of these results in forthcoming work~\cite{Woods2024b}, by developing an experimental protocol that uses $g$-factor measurements to probe the valley-phase landscape and characterize the valley-coupling parameters that determine the statistics of $E_v$. 

\emph{Acknowledgements.}---This research was sponsored in part by the Army Research Office (ARO) under Awards No.\ W911NF-17-1-0274, No.\ W911NF-22-1-0090, and No.\ W911NF-23-1-0110. 
The views, conclusions, and recommendations contained in this document are those of the authors and are not necessarily endorsed nor should they be interpreted as representing the official policies, either expressed or implied, of the ARO or the U.S. Government. The U.S. Government is authorized to reproduce and distribute reprints for Government purposes notwithstanding any copyright notation herein.


%

\clearpage
\renewcommand\thefigure{S.\arabic{figure}}
\renewcommand\theequation{S.\arabic{equation}}
\setcounter{figure}{0}
\setcounter{equation}{0}
\setcounter{section}{0}
\onecolumngrid
\vspace{0.5in}
\begin{center}
\textbf{\large Supplementary Materials: $g$-factor theory of Si/SiGe quantum dots: spin-valley and giant renormalization effects}
\end{center}

\section{Derivation of the effective-mass model from a minimal tight-binding model}

In the main text, we employ an effective-mass model, where the two valleys of the Si band structure near the $Z$-point are treated as a pseudospin-$\frac{1}{2}$ degree of freedom. 
In this supplementary section, we show how this effective-mass Hamiltonian is derivable from a minimal tight-binding model.
The usefulness of relating the effective-mass to this minimal tight-binding model is that the form of the spin-orbit coupling in the tight-binding model is easily seen to be directly connected to momentum selection rules derived from atomistic tight-binding models \cite{Woods2023a}. 
The form of the inter-valley spin-orbit coupling in Eq. (\ref{Htau}) of the main text is then a result of expanding the tight-binding model around the valley minima and taking the continuum limit.
For simplicity, we do not incorporate the magnetic field within the tight-binding model, as the main point is to show how the spin-orbit term arises in the effective-mass model.

Our minimal tight-binding model is an augmented version of the tight-binding model of Boykin \textit{et. al} \cite{Boykin2004a}, where our new addition incorporates Dresselhaus spin-orbit coupling.
Similar to Ref. \cite{Losert2023}, we consider a tetragonal lattice with longitudinal and transverse lattice constants $a_l = a_0/4 = 0.136~\T{nm}$ and $a_t = a_0/\sqrt{2} = 0.384~\T{nm}$, where $a_0$ is the cubic lattice constant of Si.
We index these sites by $(i,j,m)$, where $i$, $j$, and $m$ are the indices in the $x$, $y$, and $z$ directions, respectively.
Let $\ket{i,j,m,\sigma}$ denote the state on site $(i,j,m)$ with spin $\sigma \in \{\uparrow, \downarrow\}$.
The tight-binding model then has matrix elements
\begin{equation}
    \begin{split}
        \mel{i,j,m,\sigma}{H_{TB}}{i^\prime,j^\prime,m^\prime,\sigma^\prime} =&~ 
        \delta_{i,j,\sigma}^{i^\prime, j^\prime,\sigma^\prime}
        \left(
        t_1 \delta_{m}^{m^\prime \pm 1}
        + t_2 \delta_{m}^{m^\prime \pm 2}
        - 2(t_1+t_2) \delta_{m}^{m^\prime}
        \right) \\
        &+ t_t \delta_{m,\sigma}^{m^\prime,\sigma^\prime} 
        \left(
        \delta_{i,j}^{i^\prime \pm 1,j^\prime} 
        + \delta_{i,j}^{i^\prime,j^\prime \pm 1}
        - 4 \delta_{i,j}^{i^\prime,j^\prime}
        \right) \\
        &+ V_{i,j,m} \delta_{i,j,m,\sigma}^{i^\prime,j^\prime,m^\prime,\sigma^\prime} \\
        &+ i t_{\beta}(-1)^{m} \delta_{m}^{m^\prime} 
        \left[
        \delta_{j}^{j^\prime}
        \left(
        \delta_{i}^{i^\prime + 1} 
        - \delta_{i}^{i^\prime - 1}
        \right)
        (\sigma_{x})_{\sigma,\sigma^\prime}
        -\delta_{i}^{i^\prime}
        \left(
        \delta_{j}^{j^\prime + 1} 
        - \delta_{j}^{j^\prime - 1}
        \right)
        (\sigma_{y})_{\sigma,\sigma^\prime}
        \right],
    \end{split} \label{HTB1}
\end{equation}
where $\delta$ is the Kronecker-delta function that vanishes unless the subscripts match the superscripts.
The first line in Eq. (\ref{HTB1}) contains the hoppings in the growth direction, where $t_1 = -0.60~\T{eV}$ and $t_{2} = 0.62~\T{eV}$ are the nearest-neighbor and next-nearest neighbor hopping values. 
This next-nearest hopping model along the growth direction was first used in Ref. \cite{Boykin2004a} to match the low-energy valley features of the conduction band structure of Si.\footnote{Note that the magnitudes of $t_{1}$ and $t_{2}$ used here are slightly different than those reported in Ref. \cite{Boykin2004a}, to match the locations of the valley minima $k_z = \pm k_0$ obtained in empirical tight-binding models \cite{Niquet2009}.
In addition, we note
the different sign used for $t_{1}$, as compared to Ref. \cite{Boykin2004a}, which can be viewed as a gauge transformation in which
the sign of the even-site orbitals is flipped. This gauge choice places the valley minima here at $k_z = \pm k_{0}$, whereas
the valley minima sit at $k_z = \pm k_0 \pm 2 k_{1}$ for the gauge choice of Ref. \cite{Boykin2004a}.
}
The second line in Eq. (\ref{HTB1}) contains the normal (i.e. spin-conserving) hoppings in the transverse direction. Here, $t_t = -\hbar^2/(2m_t a_t^2)$ is used to reproduce the transverse effective mass $m_t = 0.19 m_e$ at the bottom of each valley \cite{Zwanenburg2013}.
The third line in Eq. (\ref{HTB1}) accounts for the onsite potential $V$ coming from external fields and alloy-disorder. 
Finally, the last line in Eq. (\ref{HTB1}) accounts for Dresselhaus spin-orbit coupling, where the strength is parameterized by $t_\beta$.
Here, $\sigma_j$ with $j = x,y,z$ is a Pauli matrix acting in spin space.
Importantly, the sign of these hoppings flips between each layer along the growth ($z$) direction, as indicated by the $(-1)^{m}$ factor.

\begin{figure}[t]
\begin{center}
\includegraphics[width=0.7\textwidth]{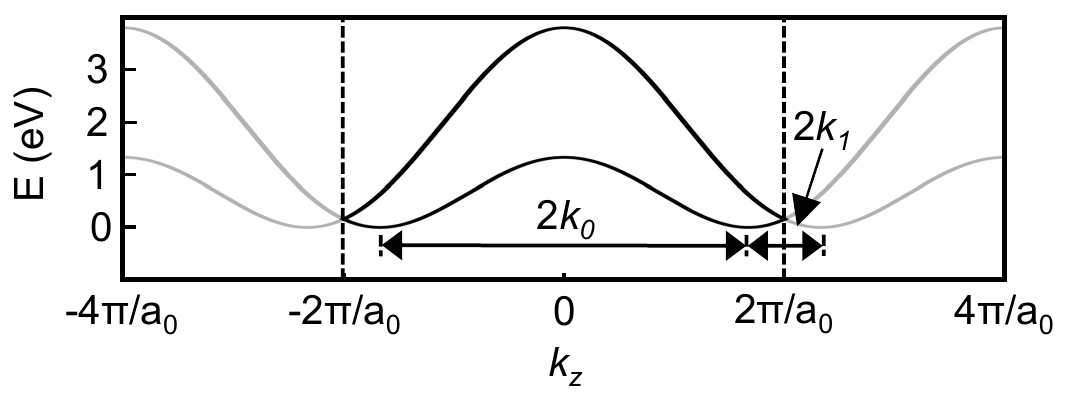}
\end{center}
\vspace{-0.5cm}
\caption{Conduction band spectrum of Si as a function of $k_z$ for $k_x\!=\!k_y\!=\! 0$.
The spectrum within the first Brillouin zone extending to $k_z = \pm 2\pi/a_0$, where $a_0$ is the cubic lattice constant of Si, is shown as black lines, while the gray lines show the spectrum repeated into the extended Brillouin zone that extends to $k_z = \pm 4\pi/a_0$. There are two valleys centered at $k_z = \pm k_0 \approx \pm 0.83 (2\pi/a_0)$. These two valleys act as a pseudospin-$\frac{1}{2}$ degree of freedom within the effective-mass model described by Eqs. (\ref{Hv}-\ref{Htau}). There are two labeled wave vectors, $2k_0$ and $2k_1 = 4\pi/a_0 - 2k_0$, that play an important role in the inter-valley coupling (see Eq. (\ref{HtauSup})).
}
\label{FIGS1}
\vspace{-1mm}
\end{figure}

To see that the tight-binding model described in real-space in Eq. (\ref{HTB1}) correctly describes the physics, it is useful to transform into momentum space.
We define momentum-space basis functions as
\begin{equation}
    \ket{\BS{k},\sigma} = 
    \frac{1}{\sqrt{N_x N_y N_z}}
    \sum_{i,j,m} 
    e^{i(k_x x_i + k_y y_j + k_z z_m)} \ket{i,j,m,\sigma},
\end{equation}
where $x_i = i a_t$, $y_j = j a_t$, $z_m = m a_l$, and $N_x$, $N_y$, and $N_z$ are the number of sites along the $x$, $y$, and $z$ axes, respectively.
Evaluating the matrix elements of the tight-binding Hamiltonian in this basis yields
\begin{equation}
    \begin{split}
        \mel{\BS{k},\sigma}{H_{TB}}{\BS{k}^\prime,\sigma^\prime} =& ~
        \delta_{\BS{k},\sigma}^{\BS{k}^\prime,\sigma^\prime}
        \Bigg[
        4t_1 \sin^2\left(\frac{k_z a_0}{8}\right)
        + 4t_2 \sin^2\left(\frac{k_z a_0}{4}\right)
        + 4t_t \sin^2\left(\frac{k_x a_t}{2}\right)
        + 4t_t \sin^2\left(\frac{k_y a_t}{2}\right)
        \Bigg]
         \\
        &+ 
        \frac{t_{\beta}}{2}
        \delta_{k_x,k_y}^{k_x^\prime,k_y^\prime}
        \delta_{|k_z - k_z^\prime|}^{4\pi/a_0}
        \Bigg[
        \sin(k_x a_t) (\sigma_{x})_{\sigma,\sigma^\prime}
        - \sin(k_y a_t) (\sigma_{y})_{\sigma,\sigma^\prime}
        \Bigg]
        + \delta_{\sigma}^{\sigma^\prime}
        \tilde{V}(\BS{k} - \BS{k}^\prime), 
    \end{split} \label{HTB2}
\end{equation}
where $\tilde{V}$ is the Fourier transform of the real-space potential,
\begin{equation}
    \tilde{V}(\BS{k}) = \frac{1}{N_x N_y N_z} \sum_{i,j,z} V_{i,j,m}
    e^{-i(k_x x_i + k_y y_j + k_z z_m)}.
\end{equation}
The spin-orbit term in Eq. (\ref{HTB2}) contains the selection rule $|k_z - k_z^\prime| = 4\pi/a_0$, which arises from the Fourier transform of the $(-1)^m$ factor in Eq. (\ref{HTB1}).
This is precisely the selection rule derived from $\T{sp}^3\T{d}^5\T{s}^*$ tight-binding models defined on a diamond lattice \cite{Woods2023a}.
Therefore, our minimal tight-binding model captures the relevant low-energy spin-orbit physics as desired.
The band structure of the tight-binding model as a function of $k_z$ for $k_x = k_y = 0$ is shown in Fig. \ref{FIGS1}.
As expected, there are two degenerate valleys at $k_z = \pm k_0$, where $k_0 \approx 0.83(2\pi/a_0)$.

The effective-mass model is derived by grouping the states around these minima, performing an expansion, and taking the continuum limit.
To do so, we first relabel our momentum basis states
\begin{equation}
    \ket{\BS{k},\sigma}_{\pm} = \ket{\BS{k} \pm k_0 \hat{\BS{z}},\sigma},
\end{equation}
where $(\pm)$ define two classes where the momentum is measured relative to $\pm k_0 \hat{\BS{z}}$.
The intra-class matrix elements are then
\begin{equation}
    \,_{\pm}\!\mel{\BS{k},\sigma}{H_{TB}}{\BS{k}^\prime,\sigma^\prime}_\pm = \delta_{\BS{k},\sigma}^{\BS{k}^\prime,\sigma^\prime}
    \left[
    \frac{\hbar^2}{2}\left( \frac{k_z^2}{m_l} + \frac{k_x^2 + k_y^2}{m_t}\right)
    \right] 
    +  \delta_{\sigma}^{\sigma^\prime}
        \tilde{V}(\BS{k} - \BS{k}^\prime)
    + \mathcal{O}(\BS{k}^3), \label{intraValleyMtxElems}
\end{equation}
and the inter-matrix elements are
\begin{equation}
    \,_{-}\!\mel{\BS{k},\sigma}{H_{TB}}{\BS{k}^\prime,\sigma^\prime}_{+} =
    \delta_{\sigma}^{\sigma^\prime}\tilde{U}(\BS{k} - \BS{k}^\prime)
    +\beta_0 \delta_{k_x,k_y}^{k_x^\prime,k_y^\prime}
        \delta_{k_z - k_z^\prime}^{-2k_1}
        \Big[
        k_x (\sigma_x)_{\sigma,\sigma^\prime}
        -k_y (\sigma_y)_{\sigma,\sigma^\prime}
        \Big]
    + \mathcal{O}(\BS{k}^3), \label{interValleyMtxElems}
\end{equation}
where $\tilde{U}(\BS{k} - \BS{k}^\prime) = \tilde{V}(\BS{k} - \BS{k}^\prime - 2k_0 \hat{\BS{z}})$ is the potential energy shifted in momentum space by $2 k_0$, $\beta_0 = t_\beta a_t/2$, and $2k_1 = 4\pi/a_0 - 2 k_0$.
Here, $2k_1$ is the inter-Brillouin zone valley separation, as indicated in Fig. \ref{FIGS1}.

The matrix elements in Eqs. (\ref{intraValleyMtxElems}, \ref{interValleyMtxElems}) are precisely (up to second-order in $\BS{k}$) those of the continuum Hamiltonian $H = H_v + H_\tau$, where $H_v$ and $H_\tau$ are the intra-valley and inter-valley terms given by
\begin{align}
    H_v &= \frac{\hbar^2}{2} 
    \left(
    \frac{\hat{k}_z^2}
    {m_l} + \frac{\hat{k}_x^2 + \hat{k}_y^2}{m_t}
    \right)
    + V(\BS{r}), \\
    H_\tau &= 
    \left[
    V(\BS{r})
    e^{-i 2k_0 z} + \beta_0 e^{i 2k_1 z}D(\hat{k}_x,\hat{k}_y)
    \right] \tau_- + h.c., \label{HtauSup}
\end{align}
where $\tau_j$ with $j \in \{x,y,z\}$ are the Pauli matrices acting in $\{+z,-z\}$-valley space, $\tau_\pm = (\tau_x \pm i\tau_y)/2$ are the valley raising and lowering operators, and $D(\hat{k}_x,\hat{k}_y) = \hat{k}_x \sigma_x - \hat{k}_y \sigma_y$ is the Dresselhaus SOC operator.
The only difference between these tight-binding and continuum representations is that, technically, the tight-binding model has $\BS{k}$ restricted to a finite range, while $\BS{k}$ is unrestricted in the continuum model.  
In practice however, this difference is unimportant, because the low-energy eigenstates are dominated by basis states $\ket{\BS{k},\sigma}_{\pm} $ with small $\BS{k}$.
This completes our derivation of the effective-mass Hamiltonian from our minimal tight-binding model.

To obtain the final form of the effective-mass Hamiltonian used in the main text, we must add in the effects of the magnetic field $\BS{B}$, which is assumed to be in-plane.
This amounts to adding a Zeeman term and altering the momentum operator according to the usual minimal coupling procedure, where $\hat{k}_z \rightarrow \hat{\pi}_z = \hat{k}_z + (\pi/\Phi_0)A_z(x,y)$.
Here, $\Phi_0 = h/2e$ is the (superconducting) magnetic flux quantum and the vector potential is given by $\BS{A}(\BS{r}) = A_z(x,y)\hat{\BS{z}}$, where $A_z(x,y) = B_x y - B_y x$ is chosen to fulfill $\BS{B} = \nabla \cross \BS{A}$.
The final form of the inter-valley Hamiltonian is then given by
\begin{equation}
    H_v = \frac{\hbar^2}{2} 
    \left(
    \frac{\hat{\pi}_z^2}
    {m_l} + \frac{\hat{k}_x^2 + \hat{k}_y^2}{m_t}
    \right)
    + V(\BS{r}) 
    + \frac{\mu_B g_0}{2} \BS{B} \cdot \BS{\sigma}, \label{HvSup}
\end{equation}
as given in Eq. (\ref{Hv}) of the main text.

We make two final comments about the model. 
First, Rashba SOC has been excluded because Dresselhaus SOC dominates over Rashba SOC in Si/SiGe heterostructures \cite{Nestoklon2006,Nestoklon2008}, which is particularly true for the WW \cite{Woods2023a}.
Second, we are confident that the spin-orbit physics is well captured in the model by comparing the resulting SOC coefficients with more-detailed empirical tight-binding models \cite{Woods2023a}. 
For example, compare $\beta_{0,0}$ in Fig. \ref{FIG4}(a) of the main text to Figs. 4 and 5 of Ref. \cite{Woods2023a}). 
Indeed, Eq. (\ref{beta00}) for $\beta_{0,0}$ allows us to fit the SOC constant $\beta_0 = 8.2~\T{meV}\cdot \T{nm}$ of our effective-mass model by matching the magnitude of the SOC coefficients of Ref. \cite{Woods2023a}.

\section{Effective Hamiltonian derivation details}

In the main text, we derived a $4 \cross 4$ effective Hamiltonian for the lowest-energy orbital of a Si/SiGe quantum dot. 
Here, we provide additional details for this calculation.

As stated in the main text, we employ the basis functions $\ket{\nu,n_x,n_y,\tau,\sigma}$, where $\nu$ is the subband index, $n_x$ and $n_y$ are the in-plane harmonic-oscillator orbital indices, $\tau \in \{+z,-z\}$ is the valley index, and $\sigma \in \{\uparrow,\downarrow\}$ is the spin index.
The real-space components of these basis functions take the form $\bra{\BS{r}}\ket{\nu,n_x,n_y} = \varphi_{\nu}(z) \chi_{n_x,n_y}(x,y)$, where $\chi_{n_x,n_y}$ is an in-plane harmonic-oscillator orbital and $\varphi_{\nu}(z) = \varphi^\T{env}_{\nu}(z)U_0(z)$ is a subband wave function with $U_0(z) \!= \!1 \!+\! (V_0/E_\lambda) \cos(G_\lambda z)$, and $\varphi^\T{env}_{\nu}$ is the subband envelope function satisfying
\begin{equation}
    \left[\frac{\hbar^2 \hat{k}_z^2}{2m_{t}} + V_l(z)\right]\varphi^\T{env}_\nu(z) = \varepsilon_\nu \varphi^\T{env}_\nu(z), \label{SubbandEqSup}
\end{equation}
where $\varepsilon_{\nu}$ is the subband energy.
Note that we can choose $\varphi_\nu(z) \in \mathbb{R}$ without loss of generality.
The Hamiltonian in this new basis is then given by 
\begin{equation}
    \begin{split}
    &H_{\BS{m},\BS{n}}^{\mu,\nu}
    =
    ~\delta_{\mu,\nu} \delta_{\BS{m},\BS{n}}
    \left[
    \varepsilon_{\nu,\BS{n}}
    + \frac{\mu_B g_0}{2} \BS{B} \cdot \BS{\sigma}
    \right] 
    + k_{z}^{\mu, \nu} \mathcal{A}_{\BS{m},\BS{n}}
    %
    + \left( 
    \left[
    \tilde{\Delta}_{\BS{m},\BS{n}}^{\mu,\nu} + \beta_{\mu, \nu}
    D_{\BS{m},\BS{n}}
    \right] \tau_-
    + h.c. \right)
    + \mathcal{V}_{\BS{m},\BS{n}}^{\mu, \nu}
    ,
    \end{split} \label{HSubband}
\end{equation}
where $H_{\BS{m}, \BS{n}}^{\mu,\nu} = \mel{\mu,\BS{m}}{H}{\nu,\BS{n}}$.
The various matrix elements in Eq.~(\ref{HSubband}) are given by 
\begin{align}
    & \varepsilon_{\nu,\BS{n}} = \varepsilon_{\nu} + \hbar \omega_t\left(n_x + n_y\right), \\
    &k_z^{\mu, \nu} = \mel{\varphi_{\mu}}{\hat{k}_z}{\varphi_{\nu}} \in i\mathbb{R}, \label{kz} \\
    &\mathcal{A}_{\BS{m},\BS{n}} = 
    \frac{\hbar^2 \pi}{m_l \Phi_0} \mel{\chi_{\BS{m}}}{A_z}{\chi_{\BS{n}}} =
    \frac{\hbar^2 \pi}{m_l \Phi_0}\left(B_x y^{\BS{m},\BS{n}} - B_y x^{\BS{m},\BS{n}}\right) \in \mathbb{R}
    ,
    \\
    &\tilde{\Delta}_{\BS{m},\BS{n}}^{\mu,\nu} = 
    \mel{\varphi_{\mu} \chi_{\BS{m}}}{V e^{-i 2k_0 z}}{\varphi_{\nu}\chi_{\BS{n}}} \in \mathbb{C}, \label{DeltaMtxElem}\\
    &\beta_{\mu, \nu} = \beta_0 \mel{\varphi_{\mu}}{e^{i 2 k_1 z}}{\varphi_{\nu}} \in \mathbb{C}, \label{betaMuNu}\\
    &D_{\BS{m},\BS{n}} = k_x^{\BS{m,\BS{n}}} \sigma_x - k_y^{\BS{m,\BS{n}}} \sigma_y, \\
    &\mathcal{V}_{\BS{m},\BS{n}}^{\mu,\nu} = 
    \mel{\varphi_{\mu} \chi_{\BS{m}}}{V_{\T{dis}}}{\varphi_{\nu}\chi_{\BS{n}}} \in \mathbb{R}, \label{VmtxElem}
\end{align}
where
\begin{align}
    x^{\BS{m}, \BS{n}} &= \delta_{m_y,n_y}\frac{\ell_t}{\sqrt{2}}\left( \sqrt{m_x}\delta_{m_x,n_x+1} + \sqrt{n_x}\delta_{m_x,n_x-1}\right) \in \mathbb{R},\\
    y^{\BS{m}, \BS{n}} &= \delta_{m_x,n_x}\frac{\ell_t}{\sqrt{2}}\left( \sqrt{m_y}\delta_{m_y,n_y+1} + \sqrt{n_y}\delta_{m_y,n_y-1}\right) \in \mathbb{R},\\
    k_x^{\BS{m}, \BS{n}} &= \delta_{m_y,n_y}\frac{i}{\sqrt{2} \ell_t}\left( \sqrt{m_x}\delta_{m_x,n_x+1} - \sqrt{n_x}\delta_{m_x,n_x-1}\right) \in i\mathbb{R},\\ 
    k_y^{\BS{m}, \BS{n}} &= \delta_{m_x,n_x}\frac{i}{\sqrt{2} \ell_t}\left( \sqrt{m_y}\delta_{m_y,n_y+1} - \sqrt{n_y}\delta_{m_y,n_y-1}\right) \in i\mathbb{R},
\end{align}
with $\ell_t = \sqrt{\hbar/( m_t \omega_t)}$ being the characteristic quantum dot size in the transverse direction. 
We also have the symmetry properties
\begin{align}
    &k_z^{\mu,\nu} = - k_z^{\nu,\mu}, \label{kzSym}  \\
    &\mathcal{A}_{\BS{m},\BS{n}} = \mathcal{A}_{\BS{n},\BS{m}}, \\
    &\tilde{\Delta}_{\BS{m},\BS{n}}^{\mu,\nu} = \tilde{\Delta}_{\BS{n},\BS{m}}^{\nu,\mu}, \\
    &\beta_{\mu,\nu} = \beta_{\nu,\mu}, \\
    &D_{\BS{m},\BS{n}} = - D_{\BS{n},\BS{m}}, \\
    &\mathcal{V}_{\BS{m},\BS{n}}^{\mu,\nu} = \mathcal{V}_{\BS{n},\BS{m}}^{\nu,\mu}, \label{VSym}
\end{align}
which are useful when performing the transformation below.

\begin{figure}[t]
\begin{center}
\includegraphics[width=0.48\textwidth]{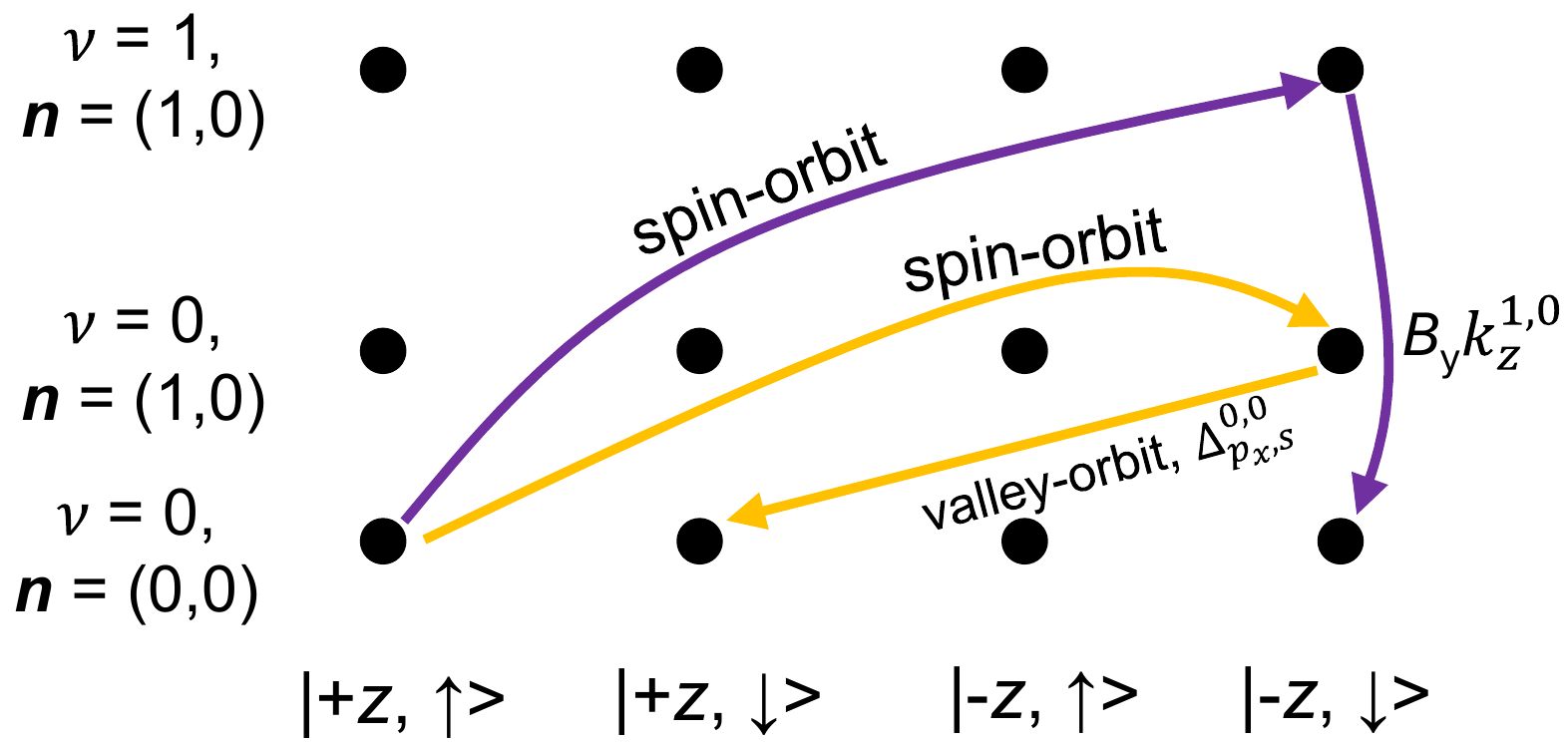}
\end{center}
\vspace{-0.5cm}
\caption{Second-order processes contributing to the effective Hamiltonian in Eq.~(\ref{HeffSup}). Columns correspond to the four combinations of the valley $\tau$ and spin $\sigma$ degrees of freedom, while the rows correspond to specified orbital quantum numbers, $(\nu,\BS{n})$.
The purple path involves the intersubband SOC and a vector potential term proportional to $B_y k_z^{1,0}$, resulting in an effective Hamiltonian term that contributes to $g_{\tau}$. 
The orange path involves the intrasubband SOC and the valley-orbit coupling $\tilde{\Delta}_{p_x,s}^{0,0} \equiv \tilde{\Delta}_{10,00}^{0,0}$ arising from alloy disorder, resulting in an effective Hamiltonian term that contributes to the spin-valley field $\BS{B}_\text{sv}$. 
}
\label{FIGS2}
\vspace{-1mm}
\end{figure}

As explained in the main text, we derive an effective low-energy Hamiltonian for the lowest-energy orbital, $(\nu,n_x,n_y) = (0,0,0)$, using a Schrieffer-Wolff transformation~\cite{Luttinger1955,Winkler2003}.
To do so, the Hamiltonian is decomposed into $H = H_0 + \tilde{H}$, where the unperturbed Hamiltonian is given by
\begin{equation}
(H_0)_{\BS{m},\BS{n}}^{\mu,\nu} = \delta_{\mu,\nu} \delta_{\BS{m},\BS{n}} \varepsilon_{\nu,\BS{n}},
\end{equation}
and the perturbing Hamiltonian $\tilde{H}$ contains the remaining terms in Eq.~(\ref{HSubband}):
\begin{equation}
   \tilde{H}_{\BS{m},\BS{n}}^{\mu,\nu}
    =
    ~\delta_{\mu,\nu} \delta_{\BS{m},\BS{n}}
    \frac{\mu_B g_0}{2} \BS{B} \cdot \BS{\sigma} 
    + k_{z}^{\mu, \nu} \mathcal{A}_{\BS{m},\BS{n}}
    %
    + \left( 
    \left[
    \tilde{\Delta}_{\BS{m},\BS{n}}^{\mu,\nu} + \beta_{\mu, \nu}
    D_{\BS{m},\BS{n}}
    \right] \tau_-
    + h.c. \right)
    + \mathcal{V}_{\BS{m},\BS{n}}^{\mu, \nu}
    . \label{HSubbandPertubation}
\end{equation}
We note that the various terms in $\tilde{H}$ are much smaller than the bare energy separations $\varepsilon_{\mu,\BS{m}} - \varepsilon_{\nu,\BS{n}}$ in $H_0$, justifying the perturbation approach. To second-order, the effective Hamiltonian is then given by 
\begin{equation}
    H_{\T{eff}} = H^{(1)} + H^{(2)}, \label{Heff3}
\end{equation}
where $H^{(1)}$ is the perturbing Hamiltonian $\tilde{H}$ projected onto the low-energy $(\nu, n_x, n_y)\! = (0,0,0)$ subspace, and $H^{(2)}$ accounts for second-order processes involving virtual states in the high-energy subspace, with  $(\nu, n_x,n_y)\! \neq (0,0,0)$.
The second-order term follows the prescription
\begin{equation}
    H^{(2)} = -{\sum_{\nu,\BS{n}}}' 
    \frac{\tilde{H}_{\BS{0},\BS{n}}^{0,\nu} \tilde{H}_{\BS{n},\BS{0}}^{\nu,0}}
    {
    \varepsilon_{\nu, \BS{n}} - \varepsilon_{0}
    }, \label{H2}
\end{equation}
where the prime mark on the summation indicates that the low-energy contribution, $(\nu,n_x,n_y) = (0,0,0)$, is not to be included.
Combining Eqs.~(\ref{HSubbandPertubation}) and (\ref{H2}), and making use of the symmetry properties in Eqs.~(\ref{kzSym})-(\ref{VSym}) yields the desired result, 
\begin{equation}
    H_{\T{eff}} = 
    ~\Delta \tau_- + \Delta^* \tau_+ 
    + \frac{\mu_B g_0}{2} \left(\BS{B} + \BS{B}_\text{sv} \tau_z\right) \cdot \BS{\sigma}
    + \frac{\mu_B}{2}\left(g_{\tau} \tau_- + g_{\tau}^* \tau_+\right)\left(B_y \sigma_x + B_x \sigma_y\right), \label{HeffSup}
\end{equation}
corresponding to Eq.~(\ref{Heff}) in the main text, where
\begin{equation}
    \Delta = \tilde{\Delta}_{\BS{0},\BS{0}}^{0,0} -2
    {\sum_{\nu,\BS{n}}}' \frac{
    \mathcal{V}^{0,\nu}_{\BS{0},\BS{n}} \tilde{\Delta}^{\nu,0}_{\BS{n},\BS{0}}
    }{\varepsilon_{\nu, \BS{n}} - \varepsilon_{0}} , \label{DeltaSup}
\end{equation}
and $g_{\tau}$ and $\BS{B}_\text{sv}$ are given by Eqs.~(\ref{gtau0}) and (\ref{Bsvx}) in the main text.
Note that the second-order term in Eq.~(\ref{DeltaSup}) gives only a small correction to the first-order term, allowing us to approximate $\Delta \approx \tilde{\Delta}_{\BS{0},\BS{0}}^{0,0}$.

Some examples of second-order processes that contribute to $g_\tau$ and $\BS{B}_\text{sv}$ are shown in Fig.~\ref{FIGS2}, where the purple and orange paths contribute to $g_\tau B_y \tau_- \sigma_x$ and $B_\text{sv}^x$, respectively.

\section{Statistics of alloy disorder matrix elements}
As explained in the main text, the valley-orbit coupling responsible for the spin-valley field $\BS{B}_\text{sv}$ arises from alloy disorder. 
It is therefore important to characterize the form and magnitude of the disorder potential $V_{\T{dis}}$ and  how they affect the statistical properties of the $\mathcal{V}_{\BS{m},\BS{n}}^{\mu,\nu}$ and $\tilde{\Delta}_{\BS{m},\BS{n}}^{\mu,\nu}$ matrix elements appearing in Eq.~(\ref{HSubband}).
In this Supplementary section, we derive these properties.

To understand the physics of alloy disorder, we consider a discretized version of the effective mass model presented in Eqs.~(\ref{Hv})-(\ref{Htau}) of the main text. Similar to Ref.~\cite{Losert2023}, we employ a tetragonal lattice with longitudinal and transverse lattice constants $a_l = a_0/4 = 0.136~\T{nm}$ and $a_t = a_0/\sqrt{2} = 0.384~\text{nm}$, where $a_0$ is the cubic lattice constant of Si. 
Also following Ref.~\cite{Losert2023}, we consider a model where the Si and Ge sites are identical, except that the Ge sites have an extra onsite energy of $E_{\T{Ge}} = 0.6~\T{eV}$, which gives the desired band offset between atomic layers with different Ge concentrations. While the actual crystal structure of Si is a diamond lattice, our choice of lattice constants, above, yields a site density that matches the atomic density of the diamond lattice of Si. We therefore expect the essential statistical properties of the alloy fluctuations to be appropriately captured in our model, as confirmed in Ref.~\cite{Losert2023} by comparing the valley splitting statistics of an effective-mass model with a large-scale $\text{sp}^{3}\T{d}^5\T{s}^*$ tight-binding model.
The main, nonfluctuating changes in the Ge concentration are assumed to occur along the growth direction $\hat z$, such that $n_{\T{Ge}} = n_{\T{Ge}}(z)$, where $n_{\T{Ge}}(z)$ is averaged over a given atomic layer in the $x$-$y$ plane. For a layer with Ge concentration $n_{\T{Ge}}(z)$, we then assume that every site is independent of the other sites, with probability $n_{\T{Ge}}(z)$ of being Ge and $(1 - n_{\T{Ge}}(z))$ of being Si. The average confinement potential of this layer is given by $E_{\T{Ge}} n_{\T{Ge}}(z)$, which we incorporate into the longitudinal potential term $V_{l}$ in Eq.~(\ref{POT}). Fluctuations about this virtual crystal potential arising from the random assignment of Ge sites are contained in the disorder potential $V_{\T{dis}}$ of Eq.~(\ref{POT}). Thus, a site at position $\BS{r}$ has a disorder potential given by
\begin{equation}
    V_{\T{dis}}(\BS{r}) = 
    \begin{cases}
        -E_{\T{Ge}}n_{\T{Ge}}(z), & \text{probability} = 1 - n_{\T{Ge}}(z) \\
        E_{\T{Ge}} \left(1 - n_{\T{Ge}}(z)\right), & \text{probability} = n_{\T{Ge}}(z)
    \end{cases}
\end{equation}
where the first and second lines correspond to Si and Ge sites, respectively.
It is easy to show that the disorder site potential satisfies
\begin{align}
    &\left< V_{\T{dis}}(\BS{r}) \right> = 0~, \label{Mean} \\
    &\left<V_{\T{dis}}(\BS{r})V_{\T{dis}}(\BS{r}^\prime)\right> = \delta_{\BS{r},\BS{r}^\prime} E_{\T{Ge}}^2 n_{\T{Ge}}(z)(1 - n_{\T{Ge}}(z)). \label{Variance}
\end{align}

Next, we consider the disorder matrix elements $\mathcal{V}_{\BS{m},\BS{n}}^{\mu,\nu}$ and $\acute{\Delta}_{\BS{m},\BS{n}}^{\mu,\nu}$, where the former is  defined in Eq.~(\ref{VmtxElem}) and appears in the intravalley Hamiltonian, while the latter is the alloy-disorder component of the normal valley coupling $\tilde{\Delta}_{\BS{m},\BS{n}}^{\mu,\nu}$, given by $\acute{\Delta}_{\BS{m},\BS{n}}^{\mu,\nu} =\mel{\varphi_{\mu} \chi_{\BS{m}}}{V_{\T{dis}}e^{-i 2k_0 z}}{\varphi_{\nu}\chi_{\BS{n}}} $.
Since we are considering the discretized version of the effective mass model on a lattice, we must use the discretized versions ($\ket{\bar{\varphi}_{\nu} \bar{\chi}_{\BS{n}}}$) of the continuum basis states, where the latter ($\ket{{\varphi}_{\nu} {\chi}_{\BS{n}}}$) are defined in Eq.~(\ref{SubbandEqSup}). 
The disorder matrix elements are then given by 
\begin{align}
    &\mathcal{V}_{\BS{m},\BS{n}}^{\mu,\nu} = \sum_{i,j} 
    V_{\T{dis}}(x_i,y_i,z_j)
    \bar{\chi}_{\BS{m}}(x_i,y_i) \bar{\chi}_{\BS{n}}(x_i,y_i)
    \bar{\varphi}_{\mu}(z_j) \bar{\varphi}_{\nu}(z_j), \\
    &\acute{\Delta}_{\BS{m},\BS{n}}^{\mu,\nu} = \sum_{i,j} 
    V_{\T{dis}}(x_i,y_i,z_j) e^{-i 2 k_0 z_j}
    \bar{\chi}_{\BS{m}}(x_i,y_i) \bar{\chi}_{\BS{n}}(x_i,y_i)
    \bar{\varphi}_{\mu}(z_j) \bar{\varphi}_{\nu}(z_j),
\end{align}
where $i$ and $j$ are the transverse and longitudinal site indices. From Eq.~(\ref{Mean}), it follows that 
\begin{equation}
    \Big<\mathcal{V}_{\BS{m},\BS{n}}^{\mu,\nu} \Big> 
    = 
    \Big< \acute{\Delta}_{\BS{m},\BS{n}}^{\mu,\nu}\Big>
    = 0. \label{Average1}
\end{equation}
Using Eq.~(\ref{Variance}), we also obtain the variance 
\begin{equation}
    \Big< \mathcal{V}_{\BS{m},\BS{n}}^{\mu,\nu} \mathcal{V}_{\BS{m}^\prime,\BS{n}^\prime}^{\mu^\prime,\nu^\prime} \Big > 
    =  
    E_{\T{Ge}}^2 
    \sum_{i} 
    \bar{\chi}_{\BS{m}\BS{m}^\prime\BS{n}\BS{n}^\prime}(x_i,y_i)
    \sum_{j} 
    \bar{\varphi}_{\mu \mu^\prime \nu \nu^\prime}(z_j)
    n_{\T{Ge}}(z_j)(1 - n_{\T{Ge}}(z_j)),
     \label{VarianceV1}
\end{equation}
where $\bar{\chi}_{\BS{m}_1\BS{m}_2\BS{m}_3\BS{m}_4}(x_i,y_i) = \prod_{p = 1}^{4} \bar{\chi}_{\BS{m}_p}(x_i,y_i)$ and $\bar{\varphi}_{\mu_1\mu_2\mu_3\mu_4}(z_j) = \prod_{p = 1}^{4} \bar{\varphi}_{\mu_p}(z_j)$.
The variance can be reformulated in terms of the continuum basis functions, using the correspondence
\begin{align}
    &\bar{\varphi}_{\nu}(z_j) \approx \sqrt{a_l} \varphi_{\nu}(z_j), \label{Contin1}\\
    &\bar{\chi}_{\BS{n}}(x_i,y_i) \approx a_t \chi_{\BS{n}}(x_i,y_i) \label{Contin2},
\end{align}
which becomes exact as $a_l,a_t \rightarrow 0$. Plugging Eqs.~(\ref{Contin1}) and (\ref{Contin2}) into
Eq.~(\ref{VarianceV1}) and approximating the sums as integrals, we find 
\begin{equation}
    \Big< \mathcal{V}_{\BS{m},\BS{n}}^{\mu,\nu} \mathcal{V}_{\BS{m}^\prime,\BS{n}^\prime}^{\mu^\prime,\nu^\prime} \Big > 
    = 
    (E_{\T{Ge}}\sqrt{a_l}a_t)^2 \Sigma^{(l)}_{\mu \nu \mu^\prime \nu^\prime}
    \Sigma^{(t)}_{\BS{m} \BS{n} \BS{m}^\prime \BS{n}^\prime}, \label{VVariance2}
\end{equation}
where the longitudinal and transverse covariance tensors are given by
\begin{align}
    \Sigma^{(l)}_{\mu \nu \mu^\prime \nu^\prime} =& 
    \int \varphi_{\mu}(z) \varphi_{\nu}(z) \varphi_{\mu^\prime}(z) \varphi_{\nu^\prime}(z) n_{\T{Ge}}(z)\left(1 - n_{\T{Ge}}(z)\right) \, dz, \label{SigmaLong}\\
    \Sigma^{(t)}_{\BS{m} \BS{n} \BS{m}^\prime \BS{n}^\prime} =& 
    \int 
    \chi_{\BS{m}}(x,y)
    \chi_{\BS{n}}(x,y)
    \chi_{\BS{m}^\prime}(x,y)
    \chi_{\BS{n}^\prime}(x,y) \, dxdy. \label{SigmaTrans}
\end{align}
Note that the elements of the covariance tensors reduce in magnitude as the basis states further spread out in space. This reduction is a consequence of averaging over more atoms. 
Moreover, Eq.~(\ref{SigmaLong}) implies that, when $n_\text{Ge}<0.5$, the longitudinal covariance tensor increases in magnitude when the subband wave functions overlap with regions of higher Ge concentration.
Also note that the covariance tensor elements are invariant to the order of the subscripts. One convenient property of the covariance tensors is that they make no mention of the discretized lattice. Therefore, we are free to use the continuum effective mass model or a discretized effective mass model with arbitrarily small lattice constants. 
For the geometry considered here, the elements of $\Sigma^{(t)}$ may be analytically calculated because $\{\chi_{\BS{n}}\}$ are harmonic-oscillator orbitals. For example, letting $\BS{m}\!=\!\BS{n}\!=\!\BS{m}^\prime\!=\!\BS{n}^\prime\!=\!\BS{0}$, we find 
\begin{equation}
    \Sigma^{(t)}_{\BS{0} \BS{0} \BS{0} \BS{0}} =  
    \frac{m_t \omega_t}{2 \pi \hbar}
    =\frac{1}{2 \pi \ell_t^2}.
\end{equation}
where $\ell_t = \sqrt{\hbar/(m_t \omega_t)}$ is the transverse localization length.

The statistics of $\acute{\Delta}_{\BS{m},\BS{n}}^{\mu,\nu}$ are easiest to understand by decomposing these elements as 
\begin{equation}
    \acute{\Delta}_{\BS{m},\BS{n}}^{\mu,\nu} = 
    R_{\BS{m},\BS{n}}^{\mu,\nu} + i I_{\BS{m},\BS{n}}^{\mu,\nu}, \label{DRI}
\end{equation}
where $R_{\BS{m},\BS{n}}^{\mu,\nu}, I_{\BS{m},\BS{n}}^{\mu,\nu} \in \mathbb{R}$. We then find that
\begin{align}
    \Big< R_{\BS{m},\BS{n}}^{\mu,\nu} R_{\BS{m}^\prime,\BS{n}^\prime}^{\mu^\prime,\nu^\prime} \Big > 
    &=  
    \frac{1}{2}
    (E_{\T{Ge}} a_t)^2 
    \Sigma^{(t)}_{\BS{m} \BS{n} \BS{m}^\prime \BS{n}^\prime}
    \sum_{j} 
    \left[1 + \cos(4 k_0 z_j)\right]
    \bar{\varphi}_{\mu}(z_j)
    \bar{\varphi}_{\mu^\prime}(z_j)
    \bar{\varphi}_{\nu}(z_j)
    \bar{\varphi}_{\nu^\prime}(z_j) 
    n_{\T{Ge}}(z_j)\left(1 - n_{\T{Ge}}(z_j\right)
    , \label{VarianceR1} \\
    \Big< I_{\BS{m},\BS{n}}^{\mu,\nu} I_{\BS{m}^\prime,\BS{n}^\prime}^{\mu^\prime,\nu^\prime} \Big > 
    &=  
    \frac{1}{2}
    (E_{\T{Ge}} a_t)^2 
    \Sigma^{(t)}_{\BS{m} \BS{n} \BS{m}^\prime \BS{n}^\prime}
    \sum_{j} 
    \left[1 - \cos(4 k_0 z_j)\right]
    \bar{\varphi}_{\mu}(z_j)
    \bar{\varphi}_{\mu^\prime}(z_j)
    \bar{\varphi}_{\nu}(z_j)
    \bar{\varphi}_{\nu^\prime}(z_j)
    n_{\T{Ge}}(z_j)\left(1 - n_{\T{Ge}}(z_j\right)
    , \label{VarianceI1} \\
    \Big< R_{\BS{m},\BS{n}}^{\mu,\nu} I_{\BS{m}^\prime,\BS{n}^\prime}^{\mu^\prime,\nu^\prime} \Big > 
    &=  
    \frac{1}{2}
    (E_{\T{Ge}} a_t)^2 
    \Sigma^{(t)}_{\BS{m} \BS{n} \BS{m}^\prime \BS{n}^\prime}
    \sum_{j} 
    \sin(4 k_0 z_j)
    \bar{\varphi}_{\mu}(z_j)
    \bar{\varphi}_{\mu^\prime}(z_j)
    \bar{\varphi}_{\nu}(z_j)
    \bar{\varphi}_{\nu^\prime}(z_j)
    n_{\T{Ge}}(z_j)\left(1 - n_{\T{Ge}}(z_j\right).
    \label{VarianceRI1} 
\end{align}
Here we notice that the sine and cosine terms in Eqs.~(\ref{VarianceR1})-(\ref{VarianceRI1}) oscillate rapidly in space. Hence, these terms tend towards zero when averaged over the wave function envelopes. For simplicity, we ignore these contributions, leading to the simplified expressions
\begin{align}
    \Big< R_{\BS{m},\BS{n}}^{\mu,\nu} R_{\BS{m}^\prime,\BS{n}^\prime}^{\mu^\prime,\nu^\prime} \Big > 
    &\approx 
    \Big< I_{\BS{m},\BS{n}}^{\mu,\nu} I_{\BS{m}^\prime,\BS{n}^\prime}^{\mu^\prime,\nu^\prime} \Big >
    \approx
    \frac{1}{2}(E_{\T{Ge}}\sqrt{a_l}a_t)^2 \Sigma^{(l)}_{\mu \nu \mu^\prime \nu^\prime}
    \Sigma^{(t)}_{\BS{m} \BS{n} \BS{m}^\prime \BS{n}^\prime}, \label{RRSimp} \\
    \Big< R_{\BS{m},\BS{n}}^{\mu,\nu} I_{\BS{m}^\prime,\BS{n}^\prime}^{\mu^\prime,\nu^\prime} \Big > 
    &\approx 0. \label{RISimp}
\end{align}
From Eq.~(\ref{RISimp}), we see that the real and imaginary components of the matrix elements of $\acute{\Delta}$ are uncorrelated. Furthermore, we see from Eqs.~(\ref{VVariance2}) and (\ref{RRSimp}) that the matrix elements of the intravalley $\mathcal{V}$ and intervalley $\acute{\Delta}$ are on the same order. Note that fast oscillations also occur when calculating the correlations between $\mathcal{V}$ and $\acute{\Delta}$ matrix elements, leading to 
\begin{equation}
    \Big< \mathcal{V}_{\BS{m},\BS{n}}^{\mu,\nu} R_{\BS{m}^\prime,\BS{n}^\prime}^{\mu^\prime,\nu^\prime} \Big >
    \approx
    \Big< \mathcal{V}_{\BS{m},\BS{n}}^{\mu,\nu} I_{\BS{m}^\prime,\BS{n}^\prime}^{\mu^\prime,\nu^\prime} \Big >
    \approx 0.
\end{equation}
In other words, the matrix elements of $\mathcal{V}$ and $\acute{\Delta}$ are uncorrelated with one another.

\section{Generating valley splitting landscapes}
In Fig.~\ref{FIG1}(b) of the main text, we presented an example valley splitting $E_v$ landscape to illustrate the presence of points of vanishing valley splitting.
Here, we explain how such landscapes are numerically generated.

To generate the valley splitting $E_v = 2|\Delta| \approx 2|\tilde{\Delta}_{\BS{0},\BS{0}}^{0,0}|$ as a function of position, we need to take into account the correlations of $\tilde{\Delta}_{\BS{0},\BS{0}}^{0,0}$ between different spatial locations where the quantum dot is centered. 
We therefore extend the definition of valley coupling matrix elements $\tilde{\Delta}_{\BS{m},\BS{n}}^{\mu,\nu}$ to become functions of position, 
\begin{equation}
    \tilde{\Delta}_{\BS{m},\BS{n}}^{\mu,\nu} \rightarrow \tilde{\Delta}_{\BS{m},\BS{n}}^{\mu,\nu}(x,y) = \int \varphi_\mu^*(z^\prime) \chi_{\BS{m}}^*(x^\prime - x, y^\prime - y)
    V(\BS{r}^\prime)e^{-i 2k_0 z^\prime} \varphi_{\nu}(z^\prime) \chi_{\BS{n}}(x^\prime - x, y^\prime - y) \, d\BS{r}^\prime,
\end{equation}
where the centers of the in-plane harmonic-oscillator orbitals have been shifted to $(x,y)$.
The correlation functions in Eqs.~(\ref{RRSimp}) and (\ref{RISimp}) are likewise generalized to include the coordinates of the quantum dot center,
\begin{align}
    \Big< 
    R_{\BS{m},\BS{n}}^{\mu,\nu}(x,y)
    R_{\BS{m}^\prime,\BS{n}^\prime}^{\mu^\prime,\nu^\prime} (x^\prime,y^\prime)
    \Big > 
    &= 
    \Big< 
    I_{\BS{m},\BS{n}}^{\mu,\nu} (x,y)
    I_{\BS{m}^\prime,\BS{n}^\prime}^{\mu^\prime,\nu^\prime} (x^\prime,y^\prime)
    \Big >
    =
    \frac{1}{2}(E_{\T{Ge}}\sqrt{a_l}a_t)^2 \Sigma^{(l)}_{\mu \nu \mu^\prime \nu^\prime}
    \Sigma^{(t)}_{\BS{m} \BS{n} \BS{m}^\prime \BS{n}^\prime}(x,y,x^\prime,y^\prime), \label{RRSimp2} \\
    \Big< R_{\BS{m},\BS{n}}^{\mu,\nu} (x,y)
    I_{\BS{m}^\prime,\BS{n}^\prime}^{\mu^\prime,\nu^\prime} (x^\prime,y^\prime)
    \Big > 
    &= 0. \label{RISimp2}
\end{align}
where the generalized transverse covariance tensor elements are given by
\begin{equation}
    \Sigma^{(t)}_{\BS{m} \BS{n} \BS{m}^\prime \BS{n}^\prime}(x,y,x^\prime,y^\prime) =
    \int 
    \chi_{\BS{m}}(x^{\prime\prime} - x,y^{\prime\prime} - y)
    \chi_{\BS{n}}(x^{\prime\prime}-x,y^{\prime\prime}-y)
    \chi_{\BS{m}^\prime}(x^{\prime\prime}-x^\prime,y^{\prime\prime}-y^\prime)
    \chi_{\BS{n}^\prime}(x^{\prime\prime}-x^\prime,y^{\prime\prime}-y^\prime) \, dx^{\prime\prime}dy^{\prime\prime},\label{SigmaTrans2}
\end{equation}
which clearly has the translation invariance property $\Sigma^{(t)}_{\BS{m} \BS{n} \BS{m}^\prime \BS{n}^\prime}(x,y,x^\prime,y^\prime) = \Sigma^{(t)}_{\BS{m} \BS{n} \BS{m}^\prime \BS{n}^\prime}(x-x^\prime,y-y^\prime,0,0)$.
A direct calculation shows that 
\begin{equation}
    \Sigma^{(t)}_{\BS{0} \BS{0} \BS{0} \BS{0}}(x,y,0,0) = \frac{1}{2 \pi \ell_t^2}
    e^{-(x^2+y^2)/(2\ell_t^2)}.
\end{equation}
Therefore, Eq.~(\ref{RRSimp2}) for $(\mu = \nu = 0, \BS{m} = \BS{n} = \BS{0})$ is given by
\begin{equation}
    \Big< 
    R_{\BS{0},\BS{0}}^{0,0}(x,y)
    R_{\BS{0},\BS{0}}^{0,0} (0,0)
    \Big > 
    = 
    \Big< 
    I_{\BS{0},\BS{0}}^{0,0} (x,y)
    I_{\BS{0},\BS{0}}^{0,0} (0,0)
    \Big >
     = \frac{\sigma_{\Delta}^2}{2}  e^{-(x^2+y^2)/(2\ell_t^2)}, \label{RRSimp3}
\end{equation}
where $\sigma_{\Delta}^2$ is a prefactor that depends on the details of the Ge concentration profile $n_{\T{Ge}}$ and the subband wave function $\varphi_{0}$.

We then consider a two-dimensional lattice of points $\{\BS{r}_{j}^\perp\ = (x_j,y_j)\}$ for which we will generate $\Delta$ consistent with the correlations present in Eq.~(\ref{RRSimp3}). Letting $\mathcal{R}_{j} = R_{\BS{0},\BS{0}}^{0,0}(x_j,y_j)$, we can define a covariance matrix 
\begin{equation}
    \Sigma_{i,j}^{\mathcal{R}} = 
    \left<\mathcal{R}_i \mathcal{R}_j\right> = 
    \frac{\sigma_{\Delta}^2}{2}e^{-|\BS{r}_i - \BS{r}_j|^2/(2\ell_t^2)}.
\end{equation}
Being a real-symmetric matrix, $\Sigma^\mathcal{R}$ may be diagonalized by an orthogonal matrix $O$: $\tilde{\Sigma}^{\mathcal{R}} = O^T \Sigma^{\mathcal{R}} O = \T{diag}(\lambda_1,\lambda_2,\dots,\lambda_N)$, where $N$ is the total number of lattice points.
We then define new variables $\mathcal{R}_i^\prime = \sum_{j} \mathcal{R}_j O_{j,i}$, which have the property
\begin{equation}
    \left< \mathcal{R}^\prime_i \mathcal{R}_j^\prime \right> = 0, \quad
    \left< \mathcal{R}^\prime_i \mathcal{R}_j^\prime \right>   
    = \delta_{i,j} \lambda_j.
\end{equation}
Hence, each $\mathcal{R}_{j}^\prime$ variable is described by an independent normal distribution with zero mean and variance $\lambda_j$. 
The idea is to generate $\{R_j^\prime\}$ by sampling from $N$ independent normal distributions, and transforming back to the original variables via $\mathcal{R}_j = \sum_{i} \mathcal{R}_i^\prime O_{i,j}^T$. 
We can also generate random instances of $I_{\BS{0},\BS{0}}^{0,0}$ using the same procedure.
In this way, we can generate valley splitting landscapes $E_v(x,y) = 2|\Delta(x,y)|$, like the one shown in Fig.~\ref{FIG1}(b), which satisfy the spatial correlations given in Eq.~(\ref{RRSimp3}).
The method is efficient because the transformation matrix $O$ only needs to be computed once.

In Fig.~\ref{FIG1}(b) of the main text, we have adopted the parameters $\sigma_{\Delta} = 200~\mu\T{eV}$ and $\ell_t = 10~\T{nm}$, corresponding to an in-plane orbital splitting of $\hbar \omega_t = 2~\T{meV}$. 
This value of $\sigma_{\Delta}$ corresponds to a WW with $\bar{n}_\T{Ge} = 0.05$ and a harmonic confinement with  $\hbar \omega_z = 10~\T{meV}$ as a subband energy splitting.

\section{Calculating $g$ from the valley splitting landscape}
In Fig. \ref{FIG1}(c) of the main text, we plotted $g$ near a point with $E_{v} = 0$ in Fig. \ref{FIG1}(b). 
To calculate $g$, we use Eq. (\ref{gFactor}) of the main text.
Therefore, we need the valley-orbit matrix elements $\tilde{\Delta}_{10,00}^{0,0}$ and $\tilde{\Delta}_{01,00}^{0,0}$, as these determine the spin-valley field $\BS{B}_\T{sv}$ (see Eq. (\ref{Bsvx}) of the main text) and, in turn, the spin-valley energy $\varepsilon_\T{sv}$. 
One can show that 
\begin{equation}
    \partial_{x} \tilde{\Delta}_{00,00}^{0,0} = \frac{\sqrt{2}}{\ell_t} \tilde{\Delta}_{10,00}^{0,0},
    \quad\quad 
    \partial_{y} \tilde{\Delta}_{00,00}^{0,0} = \frac{\sqrt{2}}{\ell_t} \tilde{\Delta}_{01,00}^{0,0}.
\end{equation}
Therefore, we can extract $\tilde{\Delta}_{10,00}^{0,0}$ and $\tilde{\Delta}_{01,00}^{0,0}$ from the $\tilde{\Delta}_{00,00}^{0,0}$ data used to generate the $E_{v}$ landscape shown in Fig. \ref{FIG1}(b).
Finally, we use the maximum SOC coefficient $\beta_{0,0}$ value from the WW result shown in Fig. \ref{FIG4}(a), and neglect $\nu \geq 1$ contributions to the spin-valley field (i.e. only the $\nu = 0$ contribution in Eq. (\ref{Bsvx}) is considered).

\section{Explanation for the node in the intersubband spin-orbit coupling ($\beta_{1,0}$)  in a Wiggle Well}
In Eqs.~(\ref{beta00}) and (\ref{beta10}) of the main text, we provide analytical results for the intrasubband $\beta_{0,0}$ and intersubband $\beta_{1,0}$ SOC coefficients of a WW, obtained using the longitudinal harmonic-oscillator wave functions. Here, we explain why $\beta_{1,0}$ exhibits a node when $\lambda=\pi/k_1$, while $\beta_{0,0}$ does not.

\begin{figure}[t]
\begin{center}
\includegraphics[width=0.7\textwidth]{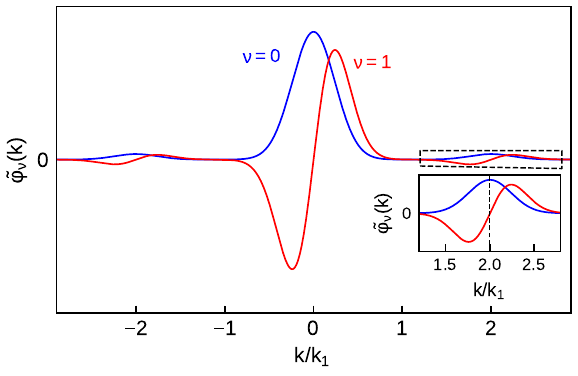}
\end{center}
\vspace{-0.5cm}
\caption{Subband wave functions $\tilde{\varphi}_{0}$ (blue) and $\tilde{\varphi}_{1}$ (red) in momentum space for a WW with $\hbar \omega_z = 20~\T{meV}$, $\lambda = \pi/k_1 =  1.57~\T{nm}$, and $V_0 = 30~\T{meV}$, corresponding to an average Ge concentration of $\bar{n}_{\T{Ge}} = 0.05$.
The Ge concentration oscillations produce satellite features near $k = \pm 2 k_1$, where the inset shows a zoom-in of the wave functions near $k = 2 k_1$.
The SOC coefficients $\beta_{0,0}$ and $\beta_{1,0}$ are both maximized near $\lambda\approx\pi/k_1$ because this maximizes the overlap integral in Eq.~(\ref{SOCcoefs2}). 
However, the intersubband SOC coefficient $\beta_{1,0}$ vanishes precisely at this resonance condition, causing a node in Fig.~\ref{FIG4}(a), because the satellite feature of $\tilde{\varphi}_{1}$ is anti-symmetric with respect to $k = 2k_1$.
}
\label{FIGS3}
\vspace{-1mm}
\end{figure}

The difference in behavior is best understood by rewriting the SOC coefficients $\beta_{\mu,\nu}$, defined in Eq.~(\ref{betaMuNu}), in momentum space:
\begin{equation}
    \beta_{\mu,\nu} = \beta_0 \int 
    \tilde{\varphi}_{\mu}^*(k + 2 k_1) \tilde{\varphi}_{\nu}(k) \, dk \label{SOCcoefs2}
\end{equation}
where 
\begin{equation}
    \tilde{\varphi}_\nu(k) \! = \! (1/\sqrt{2\pi}) \int \varphi_{\nu}(z)e^{-i k z} \, dz \label{FTSubband}
\end{equation} 
is the Fourier transform of the subband wave function $\varphi_{\nu}$.
In this way, we see that $\beta_{\mu,\nu} $ can be understood as an overlap of wave functions shifted by $2k_1$ in reciprocal space.
Plugging $\varphi_{\nu}(z) = \varphi^{\T{env}}_{\nu}(z)U_0(z)$ into Eq.~(\ref{FTSubband}), where $U_0$ is the lowest-energy $\BS{k} = 0$ Bloch function defined above Eq.~(\ref{SubbandEqSup}) and $\varphi^{\T{env}}_{\nu}$ is the subband wave function envelope defined by Eq.~(\ref{SubbandEqSup}), we find
\begin{equation}
    \tilde{\varphi}_\nu(k) = \tilde{\varphi}^{\T{env}}_\nu(k) + \frac{V_0}{2E_\lambda}
    \left( 
    \tilde{\varphi}^{\T{env}}_\nu(k+G_\lambda) 
    + \tilde{\varphi}^{\T{env}}_\nu(k-G_\lambda) 
    \right), \label{Fourier1}
\end{equation}
where 
\begin{equation}
    \tilde{\varphi}^{\T{env}}_\nu(k) \! = \! (1/\sqrt{2\pi}) \int \varphi^{\T{env}}_{\nu}(z)e^{-i k z} \, dz
\end{equation} 
is the Fourier transform of the subband wave function envelope function $\varphi^{\T{env}}_{\nu}$.
This function should be concentrated near $k = 0$, since the electron is delocalized on a length scale exceeding $\lambda$.
From Eq.~(\ref{Fourier1}), we see that the Ge concentration oscillations of a long-wavelength WW induce satellite features near $k = \pm 2k_1$, with amplitudes proportional to $V_0$. 
Such satellite peaks are indeed observed in Fig.~\ref{FIGS3} (dashed box), where $\tilde{\varphi}_0$ and $\tilde{\varphi}_1$ are shown for a WW with $\lambda = \pi/k_1$.
Since the confinement potential is taken to be symmetric in the current calculations, we obtain subband wave function envelopes that are symmetric and anti-symmetric in $k$ for the lowest-energy ($\tilde{\varphi}^{\T{env}}_0$) and first-excited subbands ($\tilde{\varphi}^{\T{env}}_1$),  respectively.
In turn, $\tilde{\varphi}_0$ and $\tilde{\varphi}_1$ are nearly symmetric and anti-symmetric about $k = G_\lambda =  2 k_1$, respectively, as shown in the inset.
The presence of satellite peaks leads to significant overlaps in Eq.~(\ref{SOCcoefs2}), and strongly enhanced peaks in $\beta_{0,0}$ and $\beta_{1,0}$  near the resonant wavevector, defined as $G_\lambda = 2k_1$.
However, the antisymmetric nature of both the main peak and satellite peaks of the $\nu=1$ (red) wave function causes the overlaps to vanish precisely at the resonance condition, resulting in a node in $\beta_{1,0}$.
Shifting $\lambda$ away from $\pi/k_1$ displaces these small satellite peaks away from $k = 2 k_1$, which reduces the overlaps and and suppresses the peaks in $\beta_{0,0}$ and $\beta_{1,0}$.
We emphasize that although the results shown in Fig.~\ref{FIGS3} were obtained a for a symmetric confinement potential $V_l$, the same qualitative behavior can also be observed for asymmetric potentials (e.g., when a WW is immersed in an electric field along the growth direction).
This is because the spectral weight of $|\tilde{\varphi}_\nu(k)|$ is always symmetric about $k = 0$.


\section{Distribution of the spin-valley energy $\varepsilon_\text{sv}$}
In Fig.~\ref{FIG5}(c) of the main text, we presented the distribution of spin-valley energies $\varepsilon_\text{sv}$ for a WW.
Here, we describe the derivation of this result.

The $x$-component of the spin-valley field is given in Eq.~(\ref{Bsvx}) of the main text, and the $y$-component is obtained by making the substitution $\tilde{\Delta}_{10,00}^{\nu,0} \rightarrow - \tilde{\Delta}_{01,00}^{\nu,0}$.
Since the subband energy splitting $\varepsilon_{\nu \neq 0} - \varepsilon_{0}$ is much larger than the in-plane orbital splitting $\hbar \omega_t$, the spin-valley field is dominated by the $\nu = 0$ contribution in Eq.~(\ref{Bsvx}), allowing us to truncate the sum.
We may perform a gauge transformation to choose a convenient phase for $\beta_{0,0}$.
Here we choose $\beta_{0,0} \in \mathbb{R}^+$, without loss of generality.
The spin-valley field is then approximately given by
\begin{equation}
    B_\text{sv}^{x} = \frac{-4 \beta_{0,0}}{\mu_B g_0} \sqrt{\frac{ m_t \omega_t}{2\hbar}} 
   \frac{\mathcal{I}_{10,00}
   }
   {\hbar \omega_t
   },
   \quad    
   B_\text{sv}^{y} = \frac{4 \beta_{0,0}}{\mu_B g_0} \sqrt{\frac{ m_t \omega_t}{2\hbar}} 
   \frac{\mathcal{I}_{01,00}
   }
   {\hbar \omega_t
   },
   \quad
   B_\text{sv}^{z} = 0,
   \label{BsvSup}
\end{equation}
where $\mathcal{I}_{\BS{m},\BS{n}} \equiv  \Im[\tilde{\Delta}_{\BS{m},\BS{n}}^{0,0}]$.
The spin-valley energy is then given by
\begin{equation}
    \varepsilon_\text{sv} = 
    \frac{2\sqrt{2} \beta_{0,0}}{\sqrt{\hbar \omega_t}} \sqrt{\frac{m_t}{\hbar^2}}
    \sqrt{
    \mathcal{I}_{10,00}^2
    + \mathcal{I}_{01,00}^2 
    }. \label{epsvSup1}
\end{equation}
Since the long-wavelength WW is expected to be in the alloy-disorder dominated valley splitting regime \cite{Woods2024a,Losert2023}, $\tilde{\Delta}_{\BS{m},\BS{n}}^{\mu,\nu}$ should have no deterministic contribution, such that $\tilde{\Delta}_{\BS{m},\BS{n}}^{\mu,\nu} \approx \acute{\Delta}_{\BS{m},\BS{n}}^{\mu,\nu}$. 
Moreover, away from any atomic-step disorder at the quantum well interface, any residual deterministic valley coupling $\tilde{\Delta}_{\BS{m},\BS{n}}^{\mu,\nu} - \acute{\Delta}_{\BS{m},\BS{n}}^{\mu,\nu}$ should be non-zero only for $\BS{m} = \BS{n}$, because $V - V_{\T{dis}}$ only depends on the $z$-coordinate away from a step. 
Therefore, $\mathcal{I}_{10,00} = \Im[\acute{\Delta}_{10,00}^{0,0}] \equiv I_{10,00}^{0,0}$ and $\mathcal{I}_{01,00} = \Im[\acute{\Delta}_{01,00}^{0,0}] \equiv I_{01,00}^{0,0}$, where we have used the definition in Eq.~(\ref{DRI}). The following statistical properties then follow from Eqs.~(\ref{Average1}),  (\ref{SigmaTrans}), and (\ref{RRSimp}):
\begin{align}
    &\left<\mathcal{I}_{10,00}\right> 
    = \left<\mathcal{I}_{01,00}\right> 
    = 0, \\
    &\left<\mathcal{I}_{10,00}^2 \right> = 
    \left<\mathcal{I}_{01,00}^2 \right> 
    = \frac{1}{2}(E_{\T{Ge}}\sqrt{a_l}a_t)^2 \Sigma^{(l)}_{0000} \left(\frac{1}{4\pi \ell_t^2} \right), \label{I1000Sq}\\
    &\left<\mathcal{I}_{10,00} \mathcal{I}_{01,00}\right> = 0, \label{I1000I0100} 
\end{align}
where Eq.~(\ref{I1000I0100}) follows from $\Sigma_{10,00,01,00}^{(t)} = 0$, because its integrand is odd in $x$ and $y$ (see Eq.~(\ref{SigmaTrans})). 
The longitudinal covariance tensor element in Eq.~(\ref{I1000Sq}) is dominated by the average Ge concentration inside the WW, $\bar{n}_{\T{Ge}}$, such that 
\begin{equation}
    \Sigma_{0000}^{(l)} \approx 
 \frac{\bar{n}_{\T{Ge}}(1 - \bar{n}_{\T{Ge}})}{\sqrt{\pi}\ell_z} 
\end{equation}
where $\ell_z = \sqrt{\hbar/(2m_l \omega_z)}$ is the longitudinal width of the wave function. 
Given that $\mathcal{I}_{10,00}$ and $\mathcal{I}_{01,00}$ are uncorrelated normal variables with zero mean, it follows that $\varepsilon_\text{sv}$ follows a Rayleigh distribution, with a probability distribution function given by
\begin{equation}
    f_\text{sv}(\varepsilon_\text{sv}) = 
    \frac{\varepsilon_\text{sv}}{\sigma_\text{sv}^2}
    e^{-\varepsilon_\text{sv}^2/(2\sigma_\text{sv}^2)}, \label{fsv}
\end{equation} 
where $\sigma_\text{sv}$ is a scale factor given by
\begin{equation}
    \sigma_\text{sv} = 
    \frac{\beta_{0,0}E_{\T{Ge}}}{\sqrt{2}\sqrt{\hbar \omega_t}} 
    \sqrt{\frac{m_t}{\hbar^2}}
    \sqrt{
    \frac{a_l a_t^2}{\pi^{3/2} \ell_z \ell_t^2} \bar{n}_{\T{Ge}}(1 - \bar{n}_{\T{Ge}})
    }. \label{sigmaSV}
\end{equation}
Note that Eq.~(\ref{sigmaSV}) is actually independent of the in-plane orbital splitting $\hbar \omega_t$ since the $\ell_t^2$ factor inside the final square root cancels out the $\sqrt{\hbar \omega_t}$ factor in the denominator of the first factor.
The distribution in Eq.~(\ref{fsv}) is plotted in Fig.~\ref{FIG5}(b) of the main text for the parameters $\hbar \omega_z = 20~\T{meV}$, $\lambda = 1.57~\T{nm}$, and $\bar{n}_{\T{Ge}} = 0.05$, which give $\sigma_\text{sv} = 4.2~\mu\T{eV}$. 

\section{Distribution of $|\Delta|/\varepsilon_\text{sv}$ for a quantum dot centered on a point of vanishing valley splitting}

As stated in the main text, we are interested in the distribution of $|\Delta|/\varepsilon_\text{sv}$, when charge noise shifts a quantum dot away from the point where $\Delta = 0$ precisely. Here, we derive Eq.~(\ref{Ratio}) of the main text, which describes the distribution of $|\Delta|/\varepsilon_\text{sv}$.

To study this problem, we consider an in-plane electric field $\BS{E} = E_x \hat{\BS{x}} + E_y \hat{\BS{y}}$, where each field component is normally distributed with zero mean and standard deviation $\sigma_{E} = 20~\mu\T{eV}/\T{nm}$.
This noise level corresponds to a standard deviation of $\sigma_{\varepsilon} = 20~\mu\T{eV}$ for the detuning parameter in a double dot system with a dot separation of $d = 100~\T{nm}$ (i.e., $\sigma_E = \sigma_\varepsilon/d$).
This detuning noise level is on the high end of what has been measured for Si/SiGe quantum dots \cite{Kranz2020}, so our calculation likely slightly overestimates typical values of $|\Delta|/\varepsilon_\text{sv}$.  

The electric field shifts the dot in space, causing a change in $\Delta$. This can be captured by considering the electric-field-induced coupling between the $s$, $p_x$, and $p_y$ harmonic-oscillator orbitals, which have quantum numbers $(n_x,n_y) = (0,0)$, $(n_x,n_y) = (1,0)$, and $(n_x,n_y) = (0,1)$, respectively. 
Ignoring spin, the Hamiltonian in the subspace containing these three low-energy orbitals and the two valley states is given by
\begin{equation}
    H = 
    \begin{pmatrix}
        0 & \Delta_{s,s}^* & -E_x \ell_t/\sqrt{2} & \Delta_{p_x,s}^* & -E_y \ell_t/\sqrt{2} & \Delta_{p_y,s}^* \\
        \Delta_{s,s} & 0 & \Delta_{p_x,s} & -E_x \ell_t/\sqrt{2} & \Delta_{p_y,s} & -E_y \ell_t/\sqrt{2} \\
        -E_x \ell_t/\sqrt{2} & \Delta_{p_x,s}^* & \hbar \omega_t & \Delta_{p_x,p_x}^* & 0 & \Delta_{p_y,p_x}^* \\
        \Delta_{p_x,s} & -E_x \ell_t/\sqrt{2} & \Delta_{p_x,p_x} & \hbar \omega_t & \Delta_{p_y,p_x} & 0 \\
        -E_y \ell_t/\sqrt{2} & \Delta_{p_y,s}^* & 0 & \Delta_{p_y,p_x}^* & \hbar \omega_t & \Delta_{p_y,p_y}^* \\
        \Delta_{p_y,s} & -E_y \ell_t/\sqrt{2} & \Delta_{p_y,p_s} & 0 & \Delta_{p_y,p_y} & \hbar \omega_t
     \end{pmatrix},
\end{equation}
where we define $\Delta_{\BS{m},\BS{n}} = \tilde{\Delta}_{\BS{m},\BS{n}}^{0,0}$.
We can obtain the effective low-energy Hamiltonian for just the two valley states in the $s$-orbital subspace by applying a second-order Schrieffer-Wolff transformation, giving 
\begin{equation}
    H_{\T{eff}} \approx 
    \begin{pmatrix}
        0 & \Delta^* \\
        \Delta & 0
    \end{pmatrix},
\end{equation}
where the effective normal valley coupling is given by
\begin{equation}
    \Delta = \Delta_{s,s} + \frac{\sqrt{2}\ell_t}{\hbar \omega_t} 
    \left( E_x \Delta_{p_x,s} + E_y \Delta_{p_y,s}\right). \label{DeltaDisplacement1}
\end{equation}

If the quantum dot is located at a point of vanishing valley splitting in the absence of charge noise, we have $\Delta_{s,s} = 0$ in Eq.~(\ref{DeltaDisplacement1}). 
For any given electric field, we can also perform a coordinate transformation such that the electric field points solely in the $x^\prime$-direction, giving
\begin{equation}
    \Delta = \frac{\sqrt{2} \ell_t}{\hbar \omega_t} |\BS{E}| \Delta_{p_{x^\prime},s},
\end{equation}
where 
\begin{equation}
    \Delta_{p_{x^\prime},s} = \cos \theta \Delta_{p_x,s} + \sin\theta \Delta_{p_y,s} ,
\end{equation}
and $\theta$ is the angle between the $x^\prime$ and $x$ axes.
Importantly, since we consider a symmetric dot, $\Delta_{p_x,s}$ and $\Delta_{p_y,s}$, $\Delta_{p_{x^\prime},s}$ should have the same statistical properties as $\Delta_{p_x,s}$ and $\Delta_{p_y,s}$. 
Thus, defining $\Delta_{p_{x^\prime},s} = \mathcal{R}_{p_{x^\prime},s} + i\mathcal{I}_{p_{x^\prime},s}$, we have 
\begin{equation}
    |\Delta| = \frac{\sqrt{2} \ell_t}{\hbar \omega_t} |\BS{E}| 
    \sqrt{
    \mathcal{R}_{p_{x^\prime},s}^2 + \mathcal{I}_{p_{x^\prime},s}^2
    }. 
    \label{DM}
\end{equation}
Here, the magnitude of the electric field $|\BS{E}|$ appears rather than its individual components. 
Since $E_x$ and $E_y$ are uncorrelated Gaussian random variables, $|\BS{E}|$ follows a Rayleigh distribution with a probability distribution function given by
\begin{equation}
    f_{E}(|\BS{E}|) =  \frac{|\BS{E}|}{\sigma_E^2}
    e^{-|\BS{E}|^2/(2\sigma_E^2)}. \label{RayleighE}
\end{equation}
Following a similar logic as described for $|\Delta|$, the spin-valley energy $\varepsilon_\text{sv}$ given in Eq.~(\ref{epsvSup1}) can be rewritten as
\begin{equation}
    \varepsilon_\text{sv} = 
    \frac{2\sqrt{2} \beta_{0,0}}{\sqrt{\hbar \omega_t}} \sqrt{\frac{m_t}{\hbar^2}}
    \sqrt{
    \mathcal{I}_{p_{x^\prime},s}^2
    + \mathcal{I}_{p_{y^\prime},s}^2 
    }. \label{epsvSup2}
\end{equation}
Combining Eqs.~(\ref{DM}) and (\ref{epsvSup2}), we obtain the relation
\begin{equation}
    \frac{|\Delta|}{\varepsilon_\text{sv}} = 
    \frac{\hbar^2|\BS{E}|}{2 m_t |\beta_{0,0}| \hbar \omega_t}  
    \Xi(\tilde{\Delta}),
    \label{RatioSup}
\end{equation}
as quoted in Eq.~(\ref{Ratio}) of the main text, where 
\begin{equation}
    \Xi(\tilde{\Delta}) = 
    \frac{
    \sqrt{
    \mathcal{R}_{p_{x^\prime},s}^2 + \mathcal{I}_{p_{x^\prime},s}^2
    }
    }
    {
    \sqrt{
    \mathcal{I}_{p_{x^\prime},s}^2
    + \mathcal{I}_{p_{y^\prime},s}^2 
    }
    }. \label{Xi}
\end{equation}

Equation~(\ref{Xi}) shows that $\Xi$ is a function of three variables: $\Xi = \Xi(\mathcal{R}_{p_{x^\prime},s}, \mathcal{I}_{p_{x^\prime},s}, \mathcal{I}_{p_{y^\prime},s})$. 
Importantly, $\mathcal{R}_{p_{x^\prime},s}$, $\mathcal{I}_{p_{x^\prime},s}$, and $\mathcal{I}_{p_{y^\prime},s}$ are independent Gaussian random variables with the same variance:
\begin{equation}
    \left<
    \mathcal{R}_{p_{x^\prime},s}^2
    \right> 
    = \left<
    \mathcal{I}_{p_{x^\prime},s}^2
    \right>
    = \left<
    \mathcal{I}_{p_{y^\prime},s}^2
    \right> 
    = \frac{1}{2}(E_{\T{Ge}}\sqrt{a_l}a_t)^2 \Sigma^{(l)}_{0000} \left(\frac{1}{4\pi \ell_t^2} \right). \label{Var1}
\end{equation}
Given that $\Xi$ in Eq.~(\ref{Xi}) involves a ratio of these normal variables, the overall scale set by Eq.~(\ref{Var1}) is irrelevant. 
In other words, the variance given in Eq. (\ref{Var1}) factors out from both the numerator and denominator in Eq. (\ref{Xi}) and cancels.
Therefore, as claimed in the main text, $\Xi$ follows a universal distribution that is independent of the details of the quantum well.
The distribution of $|\Delta|/\varepsilon_\text{sv}$ then depends only on the intrasubband SOC magnitude $|\beta_{0,0}|$, in-plane orbital splitting $\hbar \omega_t$, and the strength of the charge noise parameterized by $\sigma_E$.

Finally, the distribution of $|\Delta|/\varepsilon_\text{sv}$ values shown in Fig.~\ref{FIG5}(d) of the main text is obtained by generating $10^9$ samples of $\left\{|\BS{E}|, \mathcal{R}_{p_{x^\prime},s},\mathcal{I}_{p_{x^\prime},s},\mathcal{I}_{p_{y^\prime},s}\right\}$, where $|\BS{E}|$ is sampled from the Rayleigh distribution given in Eq.~(\ref{RayleighE}) and $ \mathcal{R}_{p_{x^\prime},s}$, $\mathcal{I}_{p_{x^\prime},s}$, and $\mathcal{I}_{p_{y^\prime},s}$ are sampled from a normal distribution with zero mean and unit variance, since the variance plays no role as explained in the previous paragraph.

\end{document}